\documentclass{aa}

\usepackage{rotating}
\usepackage{graphics}
\usepackage{psfig}
\usepackage{graphicx}
\usepackage{txfonts}

\def \hi {H\,{\sc i~}}

\def\kms{km\,s$^{-1}$}
\def\deg{\hbox{$^\circ$}}
\def\arcmin{\hbox{$^\prime$}}

\def\fdg{\hbox{$.\!\!^\circ$}}
\def\farcm{\hbox{$.\mkern-4mu^\prime$}}

\begin{document}

\title{The WSRT wide-field \hi survey} 

\subtitle{II. Local Group features\thanks{Table~\ref{tab:hvcs} is only
    available in electronic form at the CDS via anonymous ftp to
    cdsarc.u-strasbg.fr (130.79.128.5) or via
    http://cdsweb.u-strasbg.fr/cgi-bin/qcat?J/A+A/vol/page }} 

\titlerunning{Local Group
\hi in the WSRT wide-field survey} \authorrunning{Braun \& Thilker}

\author{
  Robert Braun\inst{1} \and
  David A. Thilker\inst{2} 
}

\institute{
  ASTRON,
  P.O. Box 2,
  7990 AA Dwingeloo,
  The Netherlands \and
Department of Physics and Astronomy,
Johns Hopkins University, 
3400 N. Charles St., 
Baltimore MD 21218-2695, U.S.A. }

\date{Received mmddyy / Accepted mmddyy}

\offprints{R. Braun,
\email{rbraun@astron.nl}}

\abstract{ We have used the Westerbork array to carry out an unbiased
wide-field survey of \hi emission features, achieving an {\sc rms} sensitivity
of about 18~mJy/Beam at a velocity resolution of 17~\kms\ over 1800~deg$^2$
and between $-1000~<~$V$_{\rm Hel}~<~+6500$~\kms. The primary data consists of
auto-correlation spectra with an effective angular resolution of 49$^\prime$
FWHM. The survey region is centered approximately on the position of
Messier~31 and is Nyquist-sampled over 60$\times$30$^\circ$ in
R.A.$\times$Dec. In this paper we present our \hi detections at negative
velocities which could be distinguished from the Galactic foreground. Fully
29\% of the entire survey area has high velocity \hi\ emission at a column
density exceeding our 3$\sigma$ limit of about $1.5\times10^{17}$cm$^{-2}$
over 30~\kms. We tabulate the properties of 95 discrete HVC components in the
field, corresponding to more than an order of magnitude increase in number
over that known previously. The distribution of discrete object number with
flux density has an inflection point near 12~Jy-\kms\ which may correspond to
a characteristic mass and distance for this population.  A faint population of
discrete HVCs is detected in the immediate vicinity of M31 which spans a large
fraction of the M31 rotation velocity. This class of features is confined to
about 12$^\circ$ (160~kpc) projected radius of M31 and appears to be
physically associated. Independent confirmation of the features within 3\fdg5
(47~kpc) of M31 has been obtained in our Green Bank Telescope survey (Thilker
et al. \cite{thil03}). We detect a diffuse northern extension of the
Magellanic Stream (MS) from at least $\delta=+20 \rightarrow 40^\circ$, which
subsequently loops back toward the south. Recent numerical simulations had
predicted just such an MS extension corresponding to the apo-galacticon
portion of the LMC/SMC orbit at a distance of 125~kpc. A faint bridge of \hi\
emission appears to join the systemic velocities of M31 with that of M33 and
continues beyond M31 to the north-west. This may be the first detection
of \hi\ associated with the warm-hot intergalactic medium (WHIM). The
distribution of peculiar velocity \hi\ associated with M31 can be described by
a projected exponential of 25~kpc scale-length and $5\times10^{17}$cm$^{-2}$
peak column density. We present the distribution function of N$_{HI}$ in the
extended M31 environment, which agrees well with the low red-shift QSO
absorption line data over the range log(N$_{HI}$)~=~17.2 to
log(N$_{HI}$)~=~21.9. Our data extend this comparison about two orders of
magnitude lower than previously possible and provide the first image of the
Lyman limit absorption system associated with an L$_*$ galaxy.

 }

\maketitle

\section{Introduction}

The high velocity cloud (HVC) phenomenon has been under study for some
40 years, since the first detections of $\lambda$21~cm emission from
atomic hydrogen at velocities far removed from those allowed by
rotation in the Galaxy disk (Muller, Oort \& Raimond
\cite{mull63}). Several explanations have been put forth for their
interpretation, including a galactic fountain, infall of
circum-galactic gas, tidal debris from mergers and sub-galactic-mass
companions. The suggestion has been made that at least one component of
the Galactic HVCs, the so-called CHVCs (Braun \& Burton \cite{brau99},
Blitz et al. \cite{blit99}) might be the gaseous counterpart of
low-mass dark-matter satellites. A critical prediction of this scenario
(De Heij et al. \cite{dehe02c}) is that a large population of faint
CHVCs should be detected in the vicinity of M31 (at declination
+40\deg) if enough sensitivity were available. While existing 
observational data were consistent with this scenario, they were severely
limited by the modest point source sensitivity available at northern
declinations (within the Leiden/Dwingeloo Survey, Hartmann \& Burton
\cite{hart97}) which is almost an order of magnitude poorer than that
of HIPASS (Barnes et al. \cite{barn01}) in the south.

We have undertaken a moderately sensitive large-area \hi survey both to
test for the predicted population of faint CHVCs near M31 as well as to
carry out an unbiased search for \hi emission associated with
background galaxies.  We have achieved an {\sc rms} sensitivity of
about 18~mJy/Beam at a velocity resolution of 17~\kms\ over
1800~deg$^2$ and between $-1000~<~$V$_{\rm Hel}~<~+6500$~\kms. The
corresponding {\sc rms} column density sensitivity for emission filling
the 3000$\times$2800 arcsec effective beam area is about
4$\times10^{16}$cm$^{-2}$ over 17~\kms. For comparison, the HIPASS
survey has achieved an {\sc rms} of about 14~mJy/Beam at a velocity
resolution of 18~\kms, yielding a slightly superior flux
sensitivity. On the other hand, the column density sensitivity for
emission filling our larger beam exceeds that of HIPASS by almost an
order of magnitude. Since the linear {\sc FWHM} diameter of our
survey beam varies from about 10~kpc at a distance of 
0.7~Mpc to more than 1~Mpc at 75~Mpc, it is only at Local Group
distances that the condition of beam filling is likely to be achieved.
Compared to the Leiden/Dwingeloo Survey, we achieve
an order of magnitude improvement in both flux density and brightness
sensitivity.  We detect more than 100 distinct features at high
significance in each of the two velocity regimes (negative and positive
LGSR velocities).  

In a previous paper (Braun et al. \cite{brau03a}, hereafter referred to
as BTW03) we have described the survey observations and general data
reduction procedures, together with the results for our \hi detections
of external galaxies. In this paper, we briefly relate some additional
data reduction procedures in \S\,\ref{sec:observations} and present the
results in \S\,\ref{sec:results}. In addition to a large sample of
discrete HVC components we detect three diffuse complexes of peculiar
velocity gas. We discuss these results in \S\,\ref{sec:discussion} and
close with a summary in \S\,\ref{sec:summary}.

\section{Observations and Data Reduction}
\label{sec:observations}

A complete description of the observations and general methods of data
reduction has already been given in BTW03, so we will only summarize
the essential points here and indicate any departures from the methods
employed for the positive velocity portion of the survey.

\subsection{Observations}

The survey area has an extent of 60$\times$30 true degrees oriented in
$\alpha_{2000}\times\delta_{2000}$ and centered on
($\alpha_{2000},\delta_{2000}$)~=~(10\deg, 35\deg), about 5\deg\ south
of the M31 nuclear position. Auto-correlation (and cross-correlation)
spectra were acquired with one minute data averaging using the
Westerbork Synthesis Radio Telecope (WSRT) in a series of 120 RA
drift-scans separated by 15\arcmin\ in Declination, providing Nyquist
sampling of the 35\arcmin\ FWHM primary beam. Both linear polarizations
were recorded over two 20~MHz bands centered at geocentric frequencies
of 1416 and 1398~MHz with 512 uniformly weighted spectral channels per
band.  A hanning smoothing was applied after the fact to minimize the
spectral side-lobes of interference, yielding a spectral resolution of
78.125~kHz, corresponding to about 16.6 \kms\ over a heliocentric
velocity range of about $-1000 <$V$_{\rm HEL} < +6500$ \kms.

\subsection{Data Reduction}

\begin{figure*}
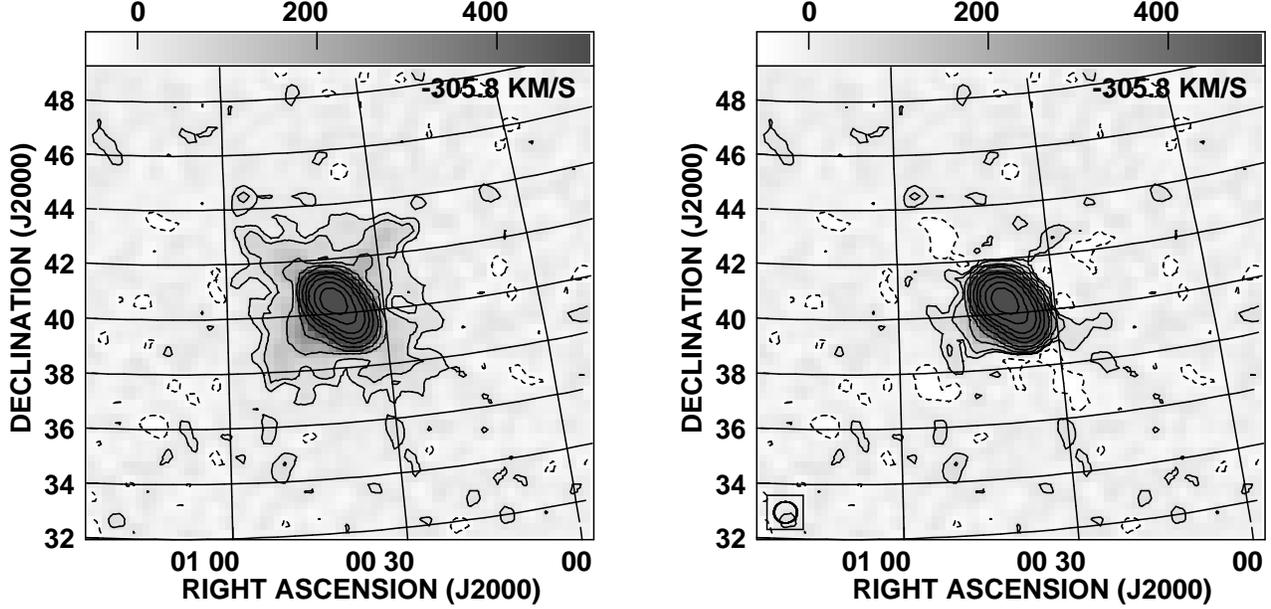

\centering
 \begin{tabular}{cc}
 \resizebox{85mm}{!}{\includegraphics{0423fg1a.ps}} &
 \resizebox{85mm}{!}{\includegraphics{0423fg1b.ps}} \\
 \end{tabular}
 \caption{Illustration of a ``dirty'' and ``clean'' channel map. \hi
 emission from a 17$\times$17\deg sub-field centered on M31 at
 V$_{\rm LGSR}$=$-$306 \kms both before (left) and after (right)
 deconvolution with the effective auto-correlation beam. The linear
 grey-scale extends from $-$50 to 500 mJy/Beam, while contours are drawn
 at $-$1, 1, 2, 5, 10, 20, 50, 100, 200, 500 and 1000 times 30
 mJy/Beam. The peak brightness in this channel is about 60
 Jy/Beam. Note the ``X''-shaped pattern of side-lobes that is visible
 out to a radius of about 3\deg in the ``dirty'' image. The dynamic
 range in the ``clean'' image is substantially improved, although there
 is a systematic over-subtraction of the near-in sidelobes, as
 discussed in the text.}
 \label{fig:dirtclean}
\end{figure*}

Data editing, calibration and gridding were employed to generate a
data-cube of the survey volume as described in BTW03. Applying the same
processing steps to a digital representation of the holographically
determined telescope primary beam allowed estimation of the effective
beam pattern of the survey. The central lobe of this effective beam can
be approximated by a Gaussian of 3020$\times$2810 arcsec at
PA~=~90\deg. The maximum near-in side-lobe levels are only about 0.25\%
of the peak response and are concentrated in an ``X'' pattern
(PA~=~$\pm$45, $\pm$135\deg) that is due to the four feed-support
legs. Beyond a radius of about 3\deg\ sidelobe levels fall below 0.05\%
even along the ``X''.  Since the telescopes of the array are
equatorially mounted, the side-lobe pattern is fixed relative to the
sky. This is an important simplifying factor for the succesful modeling
of the beam. Although moderately faint, the side-lobe response pattern
can be easily distinguished adjacent to the bright emission regions of
M31 and M33. An attempt was made to deconvolve the data-cube with the
effective beam model using the APCLN task of the Classic AIPS
package. While moderately successful, this procedure did leave some
residual artifacts at a level below about 0.1\% within a 3\deg\ radius
of the brightest emission regions. An illustration of the ``dirty'' and
``clean'' channel maps is given in Fig.~\ref{fig:dirtclean}. The sense
of the clean residual artifact noted above is a slight over-subtraction
of the near-in side-lobes. This will result in a small, systematic
underestimate of the column density of any diffuse emission features
near M31 and M33. The likely reason for this over-subtraction effect
lies in the adopted method of local spatial and velocity baselining
that was employed in calibrating the data. Extremely faint features
will not have been excluded from the sub-set of data that was used for
local spatial and velocity baselining, leading to their
suppression. This is likely to remain a challenging aspect of deep
wide-field imaging in total power.  The faint positive extensions from
the bright M31 disk seen in Fig.~\ref{fig:dirtclean} are not artifacts,
but have been confirmed as real emission features in our recent imaging
with the Green Bank Telescope (Thilker et al. \cite{thil03}). The {\sc
rms} noise level in the final data-cube was 17.4~mJy/Beam over
16.6~\kms\  at negative radial velocities ($>-1000$ \kms) which were
unconfused by emission from the Galaxy. The corresponding {\sc rms} in
\hi column density was 3.8$\times$10$^{16}$cm$^{-2}$ for emission
filling the 3020$\times$2810 arcsec beam. 

After deconvolution, a grid of smoothed data-cubes was generated at
velocity resolutions of 18, 36 and 72 \kms\ FWHM and spatial resolutions
of 48, 63 and 97\arcmin\  FWHM, to allow optimized detection of diffuse
features.  Smoothed data-cubes were used as filtering masks to isolate
regions of \hi emission from the noise background in cubes at a higher
resolution. Typical masking thresholds of 2$\sigma$ were employed in
the smoothed cubes in order to optimize signal-to-noise in the images
of integrated emission by minimizing the velocity interval of the
integration. The resulting data-cubes have been resampled in velocity
onto various systems: including heliocentric (HEL), Local Standard of
Rest (LSR), Galactic Standard of Rest (GSR) and Local Group Standard of
Rest (LGSR) defined by:

\begin{equation}
V_{\rm LSR}=V_{\rm HEL}+9\cos(l)\cos(b)+12\sin(l)\cos(b)+7\sin(b)
\end{equation}
\begin{equation}
V_{\rm GSR}=V_{\rm LSR}+0\cos(l)\cos(b)+220\sin(l)\cos(b)+0\sin(b)
\end{equation}
\begin{equation}
V_{\rm LGSR}=V_{\rm GSR}-62\cos(l)\cos(b)+40\sin(l)\cos(b)-35\sin(b)
\end{equation}

\section{Results}
\label{sec:results}

\begin{figure*}
\centering \includegraphics[width=17cm]{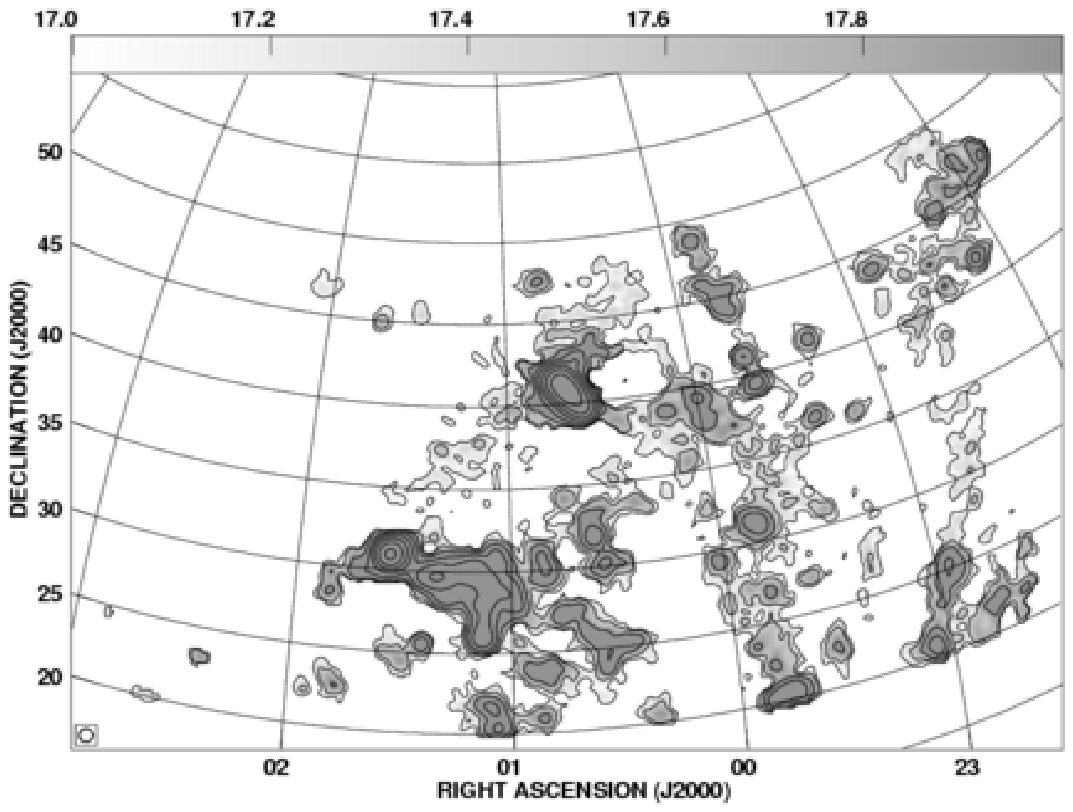}
\caption{Integrated negative velocity \hi within the survey region at
48$\arcmin$, 18~\kms\ resolution. The masked distribution of negative
velocity \hi that is distinct from Galactic IVCs has been integrated
and converted to an \hi column density under the assumption of
negligible opacity in the $\lambda21$cm line. The grey-scale varies
between log(N$_{HI}$)~=~17~--~18, for N$_{HI}$ in units of
cm$^{-2}$. Contours are drawn at log(N$_{HI}$)~=~17, 17.5, 18, $\dots$
20.5. }
\label{fig:nhiall}
\end{figure*}

The integrated \hi emission of all negative velocity features in our
survey field which could be unambiguously distinguished from the
foreground emission of our Galaxy are depicted in
Fig.~\ref{fig:nhiall}. Our criterion for distinguishing such negative
velocity features from the foreground Galactic emission was the
detection of a local peak in the emission spectrum in contrast to
merely an extended wing of emission having a spectral peak at a
velocity allowed by Galactic rotation.  While only a handful of objects
were known previously in this region; namely M31, M33, Davies' Cloud
(Davies \cite{davi75}), Wright's Cloud (Wright \cite{wrig79}), and 
9 additional high velocity clouds (De Heij et al. \cite{dehe02a}) all
tabulated in Table\ref{tab:prev}, we detect high velocity \hi emission
at a column density in excess of our 3$\sigma$ limit of about
$1.5\times10^{17}$cm$^{-2}$ from fully 29\% of the area of our 1800 deg$^2$
survey field. The detected emission is due to a mix of discrete and
diffuse components. 

\begin{table*}
\renewcommand{\thefootnote}{\thempfootnote}
\caption{ Previously detected objects in the survey field
}
\label{tab:prev}
\begin{tabular}{rccrrrr}
\hline
{Name}                   & {RA$_{2000}$}
                         & {Dec$_{2000}$}
                         & {V$_{\rm HEL}$}
                         & {V$_{\rm GSR}$}
                         & {V$_{\rm LGSR}$}
                         & {W$_{20}$}\\
{\ }                     & {$\rm(h\ m)$}
                         & {$\rm(^\circ\ ^\prime)$}
                         & {(\kms)}
                         & {(\kms)}
                         & {(\kms)}
                         & {(\kms)}\\
(1)&(2)&(3)&(4)&(5)&(6)&(7)\\
\hline

Messier 31      &00 43 &+41 16 &$-$300 &$-$122 & $-$34 &536\\
Messier 33      &01 34 &+30 40 &$-$179 & $-$44 & +37 &199\\
Davies' Cloud   &00 38 &+42 28 &$-$446 &$-$265 &$-$176 & 38\\
Wright's Cloud  &01 15 &+29 00 &$-$423 &$-$259 &$-$200 &120\\

 HVC091.8$-$35.6$-$393& 23 05& +20 43& $-$398&$-$215&$-$160& 75\\ 
CHVC107.7$-$29.7$-$429& 23 49& +31 19& $-$440&$-$247&$-$180& 61\\
CHVC108.3$-$21.2$-$402& 23 40& +39 40& $-$407&$-$208&$-$141& 50\\
CHVC118.5$-$32.6$-$386& 00 34& +30 05& $-$388&$-$223&$-$149& 40\\
 HVC118.7$-$31.1$-$135& 00 34& +31 37& $-$137&  +30& +104& 29\\
CHVC119.2$-$31.1$-$384& 00 36& +31 39& $-$386&$-$220&$-$146& 29\\
CHVC122.9$-$31.8$-$325& 00 51& +31 04& $-$326&$-$168& $-$93& 52\\
 HVC127.5$-$41.6$-$341& 01 06& +21 09& $-$339&$-$210&$-$135& 66\\
 HVC147.0$-$33.9$-$125& 02 19& +24 50& $-$120& $-$26&  +55& 107\\
\hline
\end{tabular}
\end{table*}

\subsection{A Catalog of High Velocity Clouds}

We have tabulated the properties of all local maxima with a peak column
density (at the full survey resolution) in excess of 5$\sigma$ in
Table~\ref{tab:hvcs}. Only the four previously known objects at the top of
Table\ref{tab:prev} were excluded from this tabulation. The 95 additional high
velocity features we detect in the survey field correspond to an order of
magnitude increase in number over that seen before. We have classified
each component as a CHVC (compact high velocity cloud) or an HVC (high
velocity cloud) based on the degree of isolation in an image of integrated \hi
emission made over the full velocity extent of the particular object in
question. Objects which were not connected to other features at a level
exceeding 1.5$\times$10$^{17}$cm$^{-2}$ (about 3$\sigma$) in this image were
deemed isolated and were given the CHVC designation, while those objects that
were connected to other features at this level are given the HVC designation
in the Table. The majority of the tabulated features (79 of 95) are
classified as isolated by this criterion. Previous tabulations of all-sky CHVC
and HVC populations (Putman et al. \cite{putm02}, De Heij et
al. \cite{dehe02a}) have used an isolation criterion based on a column density
of about 1.5$\times$10$^{18}$cm$^{-2}$ that was matched to the lower column
density sensitivities of the HIPASS and LDS surveys (also about
3$\sigma$). The position and size of each component has been obtained from a
2-D spatial Gaussian fit to an image of integrated emission. No correction was
made for the finite beam size of the survey. The peak column density, centroid
velocity and FWHM line-width were determined from a Gaussian fit to the single
spectrum with the maximum column density. An estimate of the total \hi flux
density for each component was made from the product of peak column density
with the spatial extent normalized by the beam area.

The columns of Table~\ref{tab:hvcs} denote the following:
\begin{itemize}
\item[] \emph{Column~1:} Running identifying number in the catalog.
\item[] \emph{Column~2:} Designation, consisting of a prefix, followed
by the Galactic longitude, Galactic latitude, and Local Standard of
Rest velocity.  The prefix is CHVC for the clouds satisfying our
isolation criteria; while HVC is used to indicate clouds connected to
more extended complexes. The longitude and latitude refer to the peak
of a spatial Gaussian fit to the integrated \hi. The velocity refers to
the peak of a Gaussian fit to the single spectrum of peak column
density.
\item[] \emph{Columns~3 and~4:} J2000 right ascension and declination
      of the position listed in column~2.
\item[] \emph{Columns~5:} Heliocentric radial velocities. 
\item[] \emph{Column~6:} Velocity FWHM of the single spectrum with peak
column density.
\item[] \emph{Column~7:} Peak column density in
      units of~$10^{18}\rm\;cm^{-2}$.
\item[] \emph{Columns~8, 9, and~10:} Angular FWHM of the major and minor
      axes, and major axis position angle. These are the results of a
spatial Gaussian fit to an image of integrated \hi, with position angle
expressed in the (RA,Dec) frame.
\item[] \emph{Column~11:} Total flux of the detection, in units 
      of $\rm Jy-km\;s^{-1}$.
\item[] \emph{Column~12:} Catalog ID's prefaced with {\tt DBB} refer to
the entries of Table~1 of De Heij et al. \cite{dehe02a}, those prefaced
with {\tt BB} to Table~1 of Braun \& Burton \cite{brau99} and those
with {\tt WW} to Table~1 of Wakker \& van Woerden \cite{wakk91}.
\end{itemize}

\begin{figure*}
\centering \includegraphics[width=17cm]{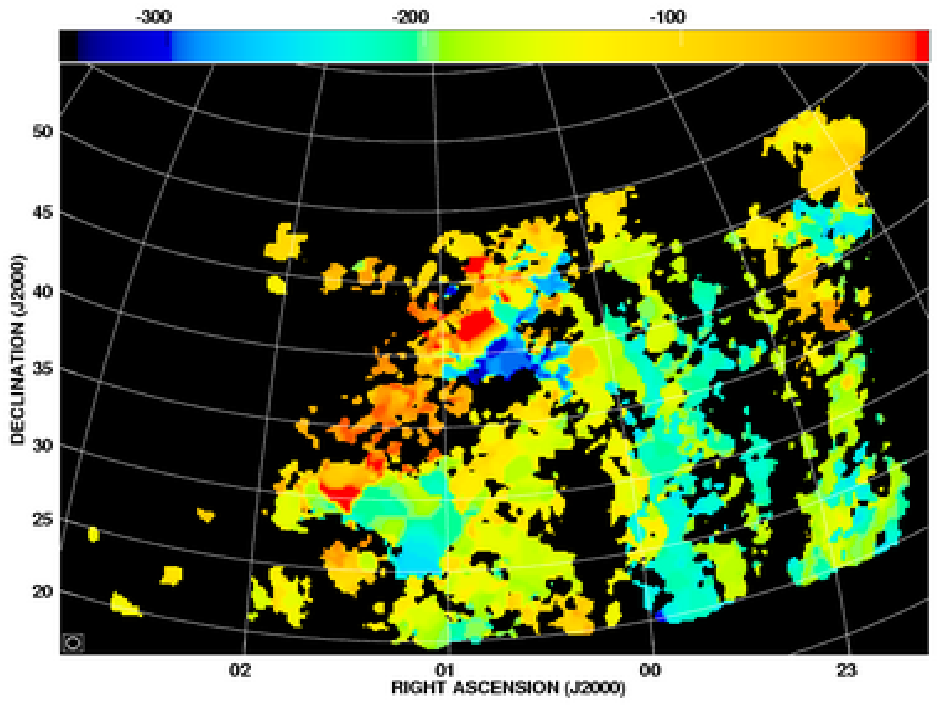}
\caption{Radial velocity in the LGSR frame of all detected \hi features
at negative velocity at 48$\arcmin$, 18~\kms\ resolution. The color
wedge varies linearly from $-$340 to $-$4~\kms and corresponds to the
velocity of the brightest emission feature along each
line-of-sight. Note the continuous variation of velocity with position
within the major diffuse features (the Magellanic Stream, Wright's Cloud
and the M31/M33 bridge) yet the large velocity offsets between them. }
\label{fig:vall}
\end{figure*}

\subsection{Kinematically Distinct Complexes of High Velocity Gas}

The line-of-sight velocities, in the LGSR system, which accompany all
detected features are depicted in Fig.~\ref{fig:vall}. Rather than
depicting the first moment of the distribution, $V_{\rm Mom}$, we have
chosen to depict the velocity of the brightest emission feature seen
along each line-of-sight, $V_{\rm Pk}$. This was done since multiple
velocity components are present in many directions. Such circumstances
are more easily recognized by the sudden changes in line-of-sight
velocity which are preserved in $V_{\rm Pk}$ while being smeared out in
$V_{\rm Mom}$.  There are several large-scale patterns in the velocity
field which permit classification of the detected features into a small
number of categories:
\begin{enumerate}
  \item A widespread component of diffuse emission is visible over much of the
western half of the survey field in the form of a filament extending
north-north-west from ($\alpha,\delta$)=(23:00,+20) to (22:30,+35)
(($l,b$)=(90,$-$36) to (93,$-$19)) and a filamentary loop extending to the
north from ($\alpha,\delta$)=(23:40,+20) to (23:40,+42) and arcing back down
to (01:00,+20) (($l,b$)=(101,$-$40) to (109,$-$19) and back to
(126,$-$43)). LGSR velocities vary smoothly along these structures from
about $-$250 to $-$150 \kms. To the north of the western-most filament are
additional diffuse features at a number of discrete velocities ranging from
$-$250 to $-$50 \kms.  Against the diffuse filamentary background are a large
number of condensations at various LGSR velocities. Some of these are
presumably associated with the diffuse features, while others appear at
distinct velocities and show no morphological connection to the diffuse
filaments. As we will show below, these diffuse filamentary features are a
previously undetected northern extension of the Magellanic Stream (MS).
\item The Local Group galaxies M31 ($V_{sys}=-$34 \kms\ in the
LGSR frame) and M33 ($V_{sys}=+37$ \kms) are visible in the central
portion of the survey field, together with a diffuse bridge of emission
that extends between them and continues to the north-west of M31 to
($\alpha,\delta$)=(00:20,+48). The radial velocity varies 
along this feature from the M31 systemic velocity near that galaxy down
to $V_{\rm LGSR}=0$ \kms\  toward M33. The bridge feature can not be traced
effectively to positive LGSR velocities due to confusing foreground
emission from the Galaxy. The full velocity interval corresponding to
rotation of M31 ($-330 > V_{\rm LGSR} > +270$ \kms) and M33 ($-85 >
V_{\rm LGSR} > +150$ \kms) is also truncated at $V_{\rm LGSR}=0$ \kms\ for the
same reason. Nonetheless, a population of faint, discrete features can
be seen near M31, spanning much of the range of M31 rotation velocities
that are unconfused. Only within a radius of about 12 degrees of M31
are such a wide variety of radial velocities detected. Many of these
discrete features have been verified in independent observations with
the Green Bank Telescope (Thilker et al. \cite{thil03}). Both the
discrete and diffuse populations near M31 appear distinct from the
Magellanic Stream features described above. The M31/M33 bridge is
offset from the nearest MS filament by about 10 degrees
on the sky and about 100 \kms\  in radial velocity.  
\item A third
component, Wright's Cloud and extensions, is seen in the south-central
portion of Fig.~\ref{fig:vall}. Although partially overlapping both the
Magellanic Stream and M33 in projection, this feature has quite
distinctive kinematics. As seen in the figure, radial velocities are
offset by more than 100 \kms\  from the M33 and to a lesser extent from
the MS. Added to this are exceptional internal kinematics which we
discuss below.
\end{enumerate}

\section{Discussion}
\label{sec:discussion}

\subsection{Discrete HVC Components}

\begin{figure*}
\centering \includegraphics[width=17cm]{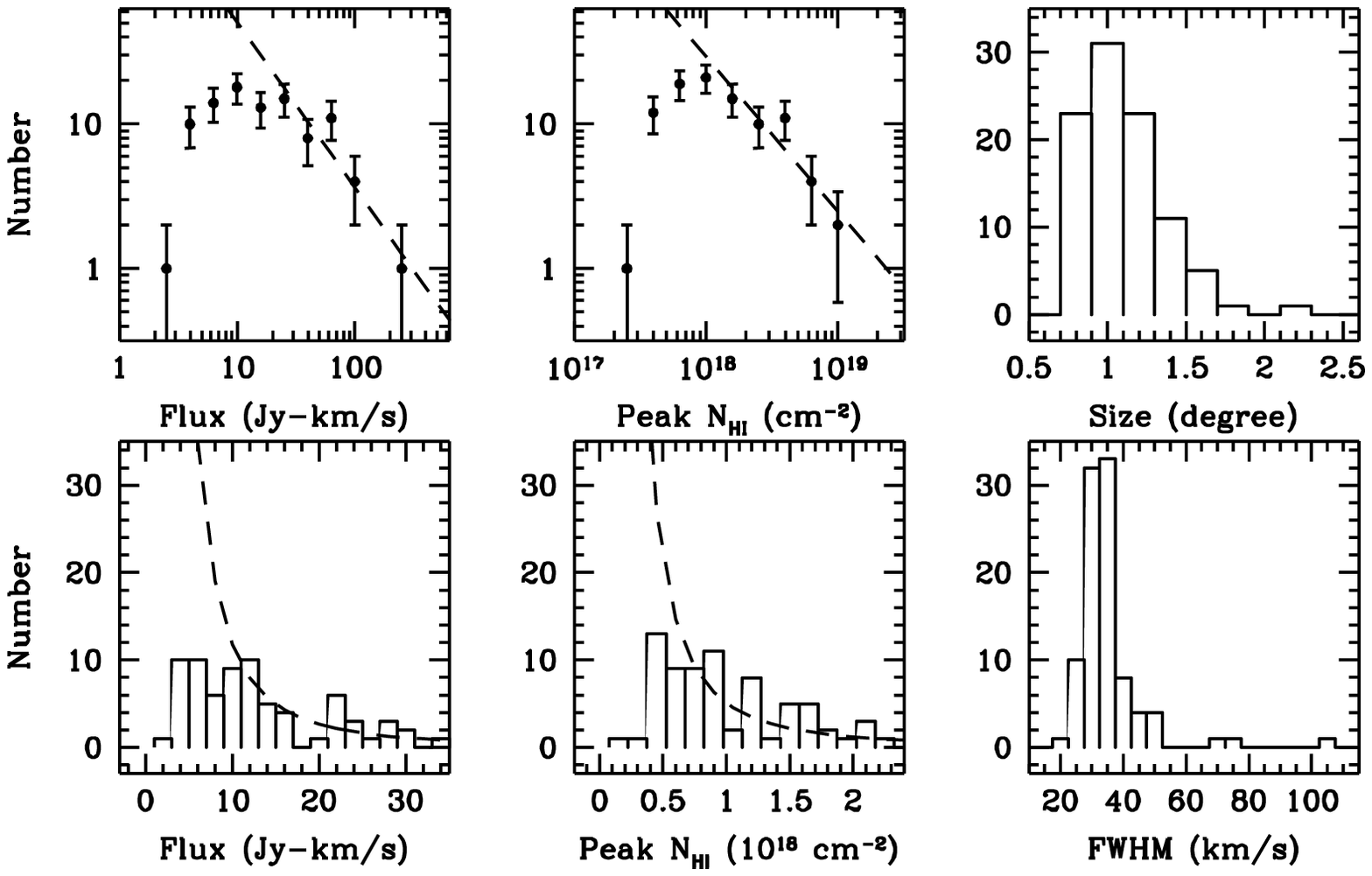}
\caption{Population properties of discrete HVC components. Histograms of
integrated flux density (left panels), peak column density (center panels),
average FWHM size (top right) and velocity FWHM (bottom right) are shown for
the entries of Table~\ref{tab:hvcs}. The linear regression solution for the
flux density (above 20 Jy-\kms) has slope $-$1.15 and for the peak column
density (above $10^{18}$cm$^{-2}$) has slope $-$1.07. The corresponding
distribution functions in linear units have power-law indices of $-$2.15 and
$-$2.07 and are overlaid in the bottom panels. }
\label{fig:hvc}
\end{figure*}

The statistical properties of the discrete high velocity components listed in
Table~\ref{tab:hvcs} are illustrated in Fig.~\ref{fig:hvc}. The distributions
of detected flux density and peak column density are both rising to lower
values. These distributions can be compared to those found previously in the
LDS survey (De Heij et al. \cite{dehe02a}) and at negative declinations in the
HIPASS survey (Putman et al. \cite{putm02}). A linear regression solution for
the logarithmically binned flux density (above 20 Jy-\kms) has slope $-$1.15
and for the peak column density (above $10^{18}$cm$^{-2}$) has slope
$-$1.07. The corresponding distribution functions in linear units can be
expressed as $f(S_{\rm HI})\propto S_{\rm HI}^{-2.15}$ and $f(N_{\rm
HI})\propto N_{\rm HI}^{-2.07}$. The linearly binned distribution
functions are also shown in the figure with the power-law solutions
overlaid. Our flux density distribution agrees well, in terms of both the
high-end slope and the apparent low-end turn-over, with that seen for the HVC
population detected in the HIPASS data. 

The apparent column density distributions, on the other hand, are much more
discrepant. Our high-end slope of $-$1.07 is much flatter than the $-$1.9 seen
in the HIPASS sample, with the apparent mode shifted to lower column
density by about an order of magnitude.  The very different beam size and
limiting column density of the two surveys (48\arcmin\ versus 16\arcmin and
$2\times 10^{17}$ versus $2\times 10^{18}$cm$^{-2}$) is likely to play an
important role in accounting for this difference. The apparent mode of the
observed FWHM size in the HIPASS survey is only about 38\arcmin, while the
mode for the WSRT wide-field survey is 60\arcmin. Both of these values
are consistent with a typical intrinsic angular FWHM size which is smaller
than the beamwidth. A more quantitative comparison is complicated by the
need to make assumptions about the intrinsic radial variation of column
density within individual features.  High resolution total power imaging of a
sample of CHVCs (Burton et al. \cite{burt01}) with the 3\farcm5 Arecibo
telescope beam has demonstrated approximately exponential column density
profiles of these objects with radius. Such profiles have the appearance of
being marginally resolved by essentially any size of telescope beam with which
they are observed. If such column density profiles are the norm, then it is
likely that both the HIPASS and current survey underestimate the intrinsic
peak column density of discrete features.

The distribution of velocity line-width is strongly peaked at a FWHM of
32~\kms, with a lower cut-off of 20~\kms\ and a sparsely populated tail
extending out to more than 100~\kms. High spatial and velocity
resolution imaging (Braun \& Burton \cite{brau00}, Burton et
al. \cite{burt01}, De Heij et al. \cite{dehe02c}) has demonstrated that
the global spectrum of such features is composed of a narrow component
due to cool (about 100~K) gas accounting for between 1 and 50\% of the
total flux, a broader component (consistent with 10$^4$K temperature)
accounting for the remainder of the \hi flux, together with the
possibility of relative motions of internal sub-structures varying from
0 to 100~\kms.  The mode of the line-width distribution, near 32~\kms,
is consistent with only modest kinematic broadening of a predominantly
warm (about 10$^4$K) \hi phase.

Both the observed size and line-width distributions agree very well
with those seen in the LDS survey (De Heij et al. \cite{dehe02a}) which
had comparable angular resolution, but much higher velocity resolution
(of 1~\kms). Extremely narrow global line profiles are thus rather
rare for high velocity \hi\ features.

The completeness of our HVC tabulation is a complicated function of position
and velocity. De Heij et al. (\cite{dehe02b}) have shown that obscuration by
Galactic \hi, coupled with the intrinsic spatial and kinematic deployment,
play a crucial role in determining which fraction of features can be detected.
At velocities free of Galactic obscuration, our {\sc rms} noise level is quite
uniform. Since both the size and line-width distributions of detected features
are strongly peaked, it is then straight-forward to estimate the typical error
in integrated flux density, $\sigma_{Typ}$~=~0.66~Jy-\kms\ (for a 60\arcmin,
32~\kms\ apparent source extent). This is a factor of about 2.2 higher than
the error for features unresolved both spatially and in velocity,
$\sigma_{Un}$~=~0.29~Jy-\kms\ (for a 48\arcmin, 17~\kms\
resolution). Simulations involving the injection of artificial sources into
total power surveys for external galaxies (eg. Rosenberg \& Schneider
\cite{rose02}) suggest that an asymptotic value of completeness is achieved
above a signal-to-noise ratio of about 8--10 in total \hi\ flux. By this
criterion, we would expect a high degree of completeness (at unconfused
velocities) for those features exceeding about $S_{Comp}$~=~6~Jy-\kms.
Interestingly, the number distribution of flux density shown in the bottom
left panel of Fig.~\ref{fig:hvc} is essentially flat between about 6 and
12~Jy-\kms and only becomes steeper at higher flux densities. The inflection
point in the distribution near S$_{Inf}$~=~12~Jy-\kms\ may be an indication
that our survey sensitivity is sufficient to detect HVC features out to some
characteristic distance beyond which their space number density diminishes
dramatically. 

As indicated above, a similar turn-over of the logarithmically binned flux
density distribution is also seen in the HIPASS sample (Putman et
al. \cite{putm02}). The HIPASS HVC flux error for unresolved sources (at
16\arcmin, 26~\kms resolution) is about 0.34~Jy-\kms, while for the typical
observed source extent of 38\arcmin\ and 32~\kms\ this increases to about
2.1~Jy-\kms. A high degree of completeness is then expected in the HIPASS HVC
sample only above about 20~Jy-\kms. The
similar turn-over of the HIPASS and WSRT wide-field distributions of
flux density below 20~Jy-\kms\ must therefore be considered fortuitous.

The velocity coverage within our survey region, $-1000~<~$V$~+6500$~\kms,
appears to have been more than sufficient to encompass all likely instances of
high velocity \hi\ associated with the Local Group. All of our HVC detections
are at negative LSR velocities, which is in keeping with the previously known
HVC properties at these Galactic longitudes (cf. De Heij et
al. (\cite{dehe02a}). Our negative velocity coverage is more extensive than
that of the LDS survey ($-450~<~$V$_{\rm LSR}~+400$~\kms) and in fact we
detect several features at more extreme LSR velocities than any which were
previously seen. The previous record holder was the object
CHVC110.6$-$08.0$-$466, discovered by Hulsbosch (\cite{huls78}). This has been
surpassed, in the general survey field (excluding the immediate vicinity of
M31), by the object CHVC093.4$-$11.9$-$485.

In the immediate vicinity of M31, an entire population of very faint
objects has been detected in our $7\times7$\deg\ Green Bank Telescope
(GBT) survey of Thilker et al. (\cite{thil03}) extending out to
$-520$~\kms. The most extreme velocity object in our tabulation which
belongs to this class, but is still distinct from the diffuse M31 disk
emission features at our survey angular resolution, is
CHVC115.4-23.1-503. With an integrated \hi\ flux of only 2.8~Jy-\kms,
this is the faintest object in our catalog. At the distance of M31,
this would correspond to an M$_{\rm HI}$= 4$\times10^5$M$_\odot$. Our
8$\sigma$ completeness limit for a spatially unresolved, 32~\kms\ source
extent corresponds to about 4.5$\times10^5$M$_\odot$ at the M31
distance.

\subsection{The Magellanic Steam}

\begin{figure*}
\centering \includegraphics[width=17cm]{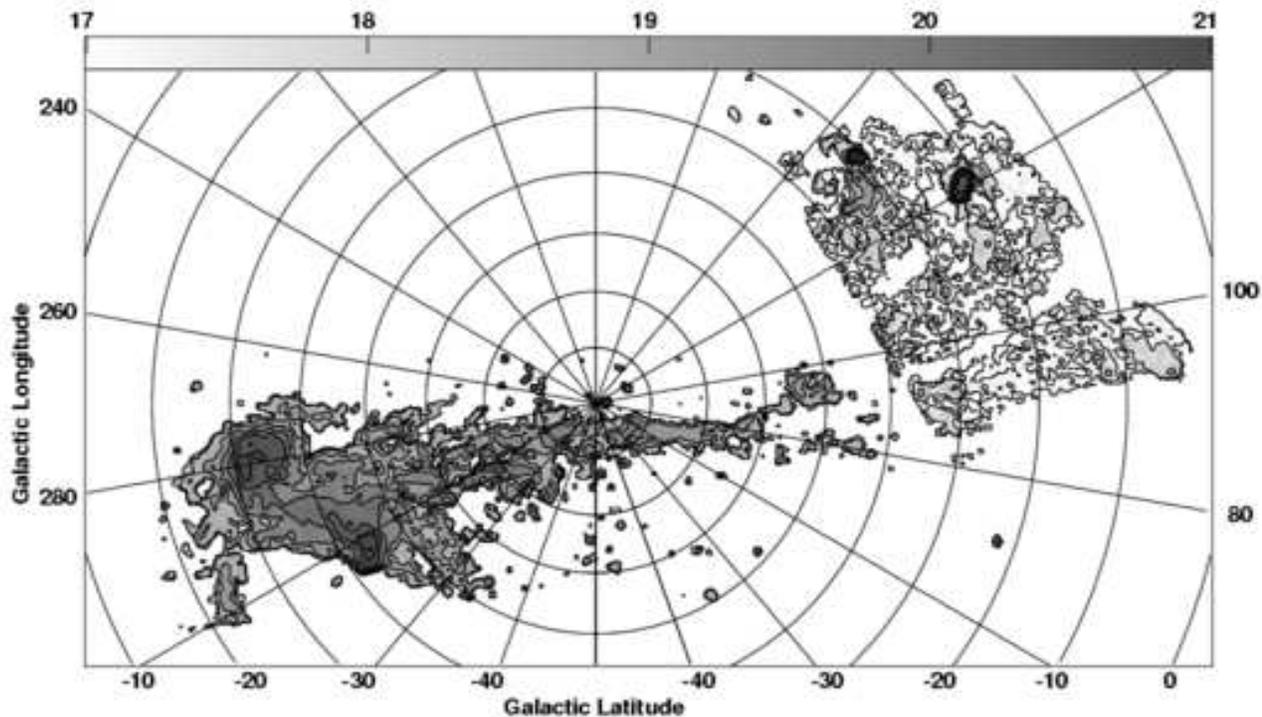}
\caption{Integrated \hi emission image highlighting the Magellanic
Clouds and Stream. The HIPASS data of Putman et al. (\cite{putm03}) is
combined with the wide-field WSRT survey data (in the region $90<l<160,
-40<b<-5$). The grey-scale varies between log(N$_{HI}$)~=~17~--~21, for
N$_{HI}$ in units of cm$^{-2}$. Contours are drawn at
log(N$_{HI}$)~=~17.5, 18.5, 19, 19.5, 20 $\dots$ 21.5. Note that the
HIPASS data is limited by sensitivity to log(N$_{HI}$)~$>$~18.3 and the
WSRT data to log(N$_{HI}$)~$>$~17.3. Kinematic evidence of association
with the Magellanic Stream only applies to a subset of the illustrated
features (see the text and Fig.\ref{fig:vall}).}
\label{fig:magstm}
\end{figure*}

An important component of the high velocity \hi gas associated with
the Galaxy is that due to the interaction of the Magellanic Clouds
with one another as well as with the Galaxy. This feature was first
assocated with the Magellanic Clouds by Mathewson et
al. (\cite{math74}) and has most recently been discussed by Putman et
al. (\cite{putm03}) based on the HIPASS survey coverage extending to
$\delta$~=~+25\deg. Modeling of the Magellanic Stream (MS)
(eg. Gardiner \& Noguchi \cite{gard96}, Gardiner \cite{gard99}) has
been moderately successful in reproducing both the distribution and
kinematics of the detected gas, but seems to require the inclusion of
some ram pressure interaction with the Galactic halo in addition to
simply gravitational interaction. Previous models which are dominated
by the effects of ram-pressure stripping (Moore \& Davis
\cite{moor94}) do not appear to reproduce the observations as
well.  The best-fitting tidal-dominated models suggest that the orbit
of the Magellanic Clouds with the Galaxy varies between about 150 and
50~kpc radius. The Clouds are currently near peri-galacticon, while
the most distant trailing portions of the Stream trace the
apo-galacticon portion of the orbit near 125~kpc, where the Clouds
were located some 0.9~Gyr ago. Ram-pressure dominated models instead
suggest that the heliocentric distance is declining along the MS, from
50~kpc at the Clouds to only some 10~kpc at the end of the tail.

We illustrate the combined spatial coverage for the MS in the HIPASS (Putman
et al. \cite{putm03}) and WSRT wide-field surveys in
Fig.~\ref{fig:magstm}. The LMC and SMC are located in the lower left portion
of the image at $(l,b)$~=~(280,$-$33) and (303,$-$44). The HIPASS coverage and
brightness sensitivity are sufficient to detect a continuous filamentary
structure beyond a bifurcation near $(l,b)$~=~(80,$-$65) out to a pair of
features near $(l,b)$~=~(80,$-$43) and (95,$-$50). With the order of magnitude
greater brightness sensitivity of the WSRT survey, the continuation of each of
these filaments can be detected over some additional 10's of degrees. The
kinematic continuity of these diffuse features is best assessed in
Fig.~\ref{fig:vall}. The westernmost filament in this $(\alpha,\delta)$
depiction is the lower filament of Fig.~\ref{fig:magstm}. It extends
continuously north to almost $\delta~=~35$\deg. We will refer to this feature
as the ``western filament''. An additional concentration near
$(\alpha,\delta)$~=~(22:20,+42) may or may not form a continuation of this
filament. A series of additional features extending in the same general
direction as the western filament has radial velocities discrepant by about
200~\kms. The relationship of these additional features to the MS is not
clear. The continuation of the upper filament of Fig.~\ref{fig:magstm} is
initially due north from $(\alpha,\delta)$~=~(23:50,+20) in
Fig.~\ref{fig:vall}. However at $\delta~=~38$\deg\ this filament is observed
to loop back to the south-east, leaving the coverage of our WSRT survey at
about $(\alpha,\delta)$~=~(01:00,+20). We will refer to this feature as the
``eastern loop''. Evidence for some of these extensions has been seen
previously in the survey of Lockman et al. (\cite{lock02}) in pointed \hi\
observations of 860 semi-random lines-of-sight. Associated O{\sc vi}
absorption from MS extensions in this region has recently been reported by
Sembach et al. (\cite{semb03}).

Both the western filament and the eastern loop are conspicuous
features of the best-fitting tidal-dominated MS models of Gardiner \&
Noguchi (\cite{gard96}) and Gardiner (\cite{gard99}). In these models,
such features are near apo-galacticon of the orbit, and correspond to
the oldest and most distant ($\sim$125~kpc) components of the
MS. Since previous observations had not yet detected these features,
they must be regarded as predictions of the Gardiner \& Noguchi model;
predictions which have been amply verified by the current survey
results.  The models suggest that the radial velocity varies smoothly
along the trailing filamentary structures of the MS and that it be
essentially single-valued from our perspective. Diffuse, contiguous
features appear to conform to these expectations. However, additional
discrete features are detected in the vicinity of, and in projection
against, the MS at various discrete velocities. There is as yet no
evidence to link them to the MS. Existing models do not predict
additional features of this type.

\subsection{Wright's Cloud}

\begin{figure}
\resizebox{\hsize}{!}{\includegraphics{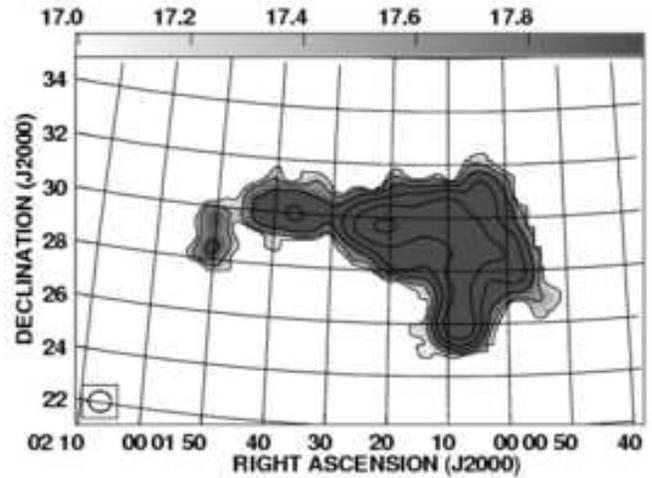}}
 \caption{Integrated \hi emission from the subset of detected features
 apparently associated with Wright's HVC.  The grey-scale varies
 between log(N$_{HI}$)~=~17~--~18, for N$_{HI}$ in units of
 cm$^{-2}$. Contours are drawn at log(N$_{HI}$)~=~17, 17.5, 18, $\dots$
 19.5. }
\label{fig:wcld}
\end{figure}

A striking example of an HVC complex which is located near the
Magellanic Stream in projection, but which is kinematically distinct is
the feature discovered by Wright (\cite{wrig79}). This complex,
centered near $(\alpha,\delta)$~=~(01:15,+29), can be seen in
Figs.~\ref{fig:nhiall} and \ref{fig:vall} sandwiched between M33 to the
east and the ``eastern loop'' feature of the MS, discussed above. The
discontinuity in radial velocity between these three different
components is apparent in Fig.~\ref{fig:vall}. The features which share
kinematic continuity with Wright's Cloud have been isolated in
Fig.~\ref{fig:wcld}. A faint, newly detected extension of the complex is
seen toward the east and south, passing across the southern half of the
M33 disk in projection. The Lockman et al. (\cite{lock02}) survey had
also seen indications for extensions of this object.

\begin{figure*}
\centering \includegraphics[width=17cm]{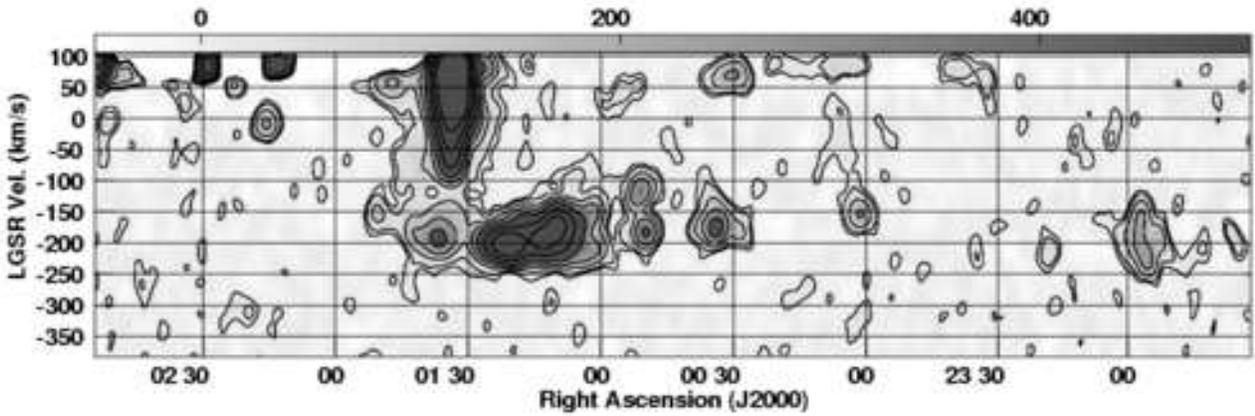}
\caption{Position-velocity diagram of the survey region at a central
 $\delta\sim30^\circ$. The linear grey-scale varies between $-$50 and
 500 mJy/Beam. Contours are drawn at 1, 2, 5, 10, 20, 50, 100 and 200
 times 20 mJy/Beam. M33 is the bright feature at
 $\alpha$=01:40, while Wright's Cloud is near $\alpha$=01:15.}
\label{fig:ravel}
\end{figure*}

\begin{figure}
\resizebox{\hsize}{!}{\includegraphics{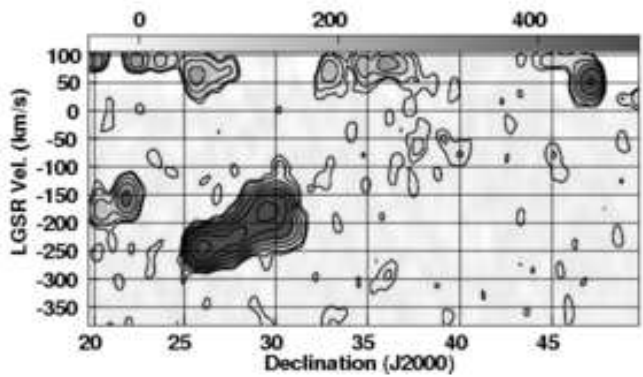}}
 \caption{Position-velocity diagram of the survey region at a central
 $\alpha$=01:07:34. The linear grey-scale varies between $-$50 and 500
 mJy/Beam.  Contours are drawn at 1, 2, 5, 10, 20, 50, 100 and 200
 times 20 mJy/Beam. Wright's Cloud extends from $\delta=26$ to
 $30^\circ$. }
\label{fig:decvel}
\end{figure}

However, the difference in $V_{\rm Pk}$ is only a part of what makes
Wright's Cloud distinctive. Position-velocity diagrams in
(SIN-)projected right ascension and declination (horizontal and
vertical in Fig.~\ref{fig:nhiall}) are shown in Figs.~\ref{fig:ravel}
and \ref{fig:decvel}. These perpendicular cross-cuts highlight the
remarkable internal kinematics of Wright's Cloud. There appear to be at
least two principal kinematic axes in this feature, oriented
approximately along PA about +75\deg and $-$10\deg. Displacement along
each of these axes is accompanied by a large difference in radial
velocity. In the north-south direction the velocity gradient is about
65~\kms\ over 3\fdg5, while in the east-west direction it amounts to
about 45~\kms\ over 3\fdg7. Multiple velocity components can be seen
along some lines-of-sight. The multi-valued velocity system adjacent to
Wright's Cloud (at $(\alpha,\delta)$~=~(00:50,+30)) appears to be a
continuation of the velocity splitting and east-west gradient seen in
the complex itself. Although this may suggest some physical association
of the feature with Wright's Cloud, it was not included in masked
version of the data which was integrated to make Fig.~\ref{fig:wcld}.
Similar velocity gradients along several distinct kinematic axes have
been detected previously in the high resolution imaging of
CHVC204+30+075 and CHVC115+13$-$275 by Braun \& Burton
(\cite{brau00}). Higher angular resolution imaging of Wright's Cloud
may supply insights into the origins of these motions. The presence of
such apparently intersecting kinematic systems argues for some form of
extreme non-equilibrium phenomenon for their origin, or chance
alignment along the line-of-sight.

The relative velocities of M33 and the Wright's Cloud extension are
also apparent in Fig.~\ref{fig:ravel}. With our modest spatial
resolution these features blend into a semi-continuous distribution,
making it difficult to distinguish one from the other. 

In the past there has been conjecture regarding the possible
association of Wright's Cloud with either M33 or the Magellanic Stream
(Wright \cite{wrig79}). Even after deep, unbiased imaging of the
broader environment of Wright's Cloud, we are still not in a good
position to assign a likely distance. The complex is moderately
distinct from M33 in position and velocity ($\Delta V\sim$200~\kms),
while a possible correspondence with the MS might be argued ($\Delta
V\sim$70~\kms). The integrated \hi\ emission from the Wright Cloud
complex is 2800~Jy-\kms, corresponding to an \hi\ mass of
4.6$\times10^8$M$_\odot$ at 840~kpc or 9.5$\times10^6$M$_\odot$ at
120~kpc (if we adopt a distance near apo-galacticon in the
tidally-dominated MS models). If the previously noted velocity
gradients are gravitational in origin we can assess the dark matter
fraction for these two assumed distances from the ratio of
R$\cdot$V$^2$/G$\cdot$M$_{HI}$. The apparent dark to \hi mass ratio
(with no inclination correction of measured velocities) varies from 13
at 840~kpc to 95 at 120~kpc. The latter possibility in particular
merits further consideration, given the similarity in apparent dark
fraction found for the discrete HVC population around M31 by Thilker
et al (\cite{thil03}). We suggest that a plausible explanation of this
feature may involve \hi\ associated with a $10^9$M$_\odot$ dark
sub-structure within the extended halo of the Galaxy.

Another notable feature which was purposely high-lighted in our choice of
cross-cut (in Fig.~\ref{fig:ravel}) is the feature we tabulate as
CHVC092.6$-$30.7$-$404 at $(\alpha,\delta)$~=~(22:56,+25). This is the
discrete HVC with the highest velocity width encountered in our sample,
with a FWHM line-width of 103~\kms. This feature is seen in projection
against the ``western filament'' of the MS, although it's velocity
width sets it apart from any other MS feature detected in the 1800
square degree survey area. If this feature were associated with this
portion of the MS, at a likely distance of 120~kpc, it would have
M$_{\rm HI}$~=~7.5$\times10^5$M$_\odot$. If the observed line-width of this
feature is gravitational then a dark:\hi\ mass ratio of about 1900 is
implied. Such a large dark fraction seems quite incredible and suggests
that other interpretations for the line-width (or the assumed distance)
might be more plausible.

\subsection{The M31/M33 Filament}

\begin{figure}
\resizebox{\hsize}{!}{\includegraphics{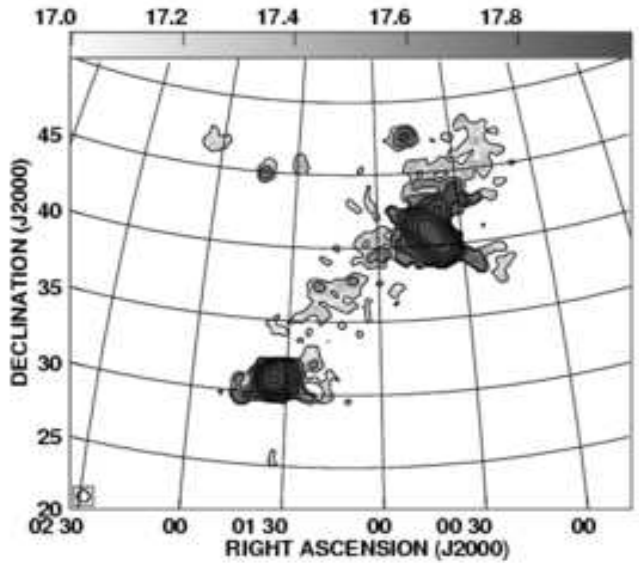}}
 \caption{Integrated \hi emission from the subset of detected features
 apparently associated with M31 and M33.  The grey-scale varies between
 log(N$_{HI}$)~=~17~--~18, for N$_{HI}$ in units of cm$^{-2}$. Contours
 are drawn at log(N$_{HI}$)~=~17, 17.5, 18, $\dots$ 20.5. }
\label{fig:m31m33}
\end{figure}

A striking feature of the low column density \hi\ sky is the apparent
bridge connecting the systemic velocities of M31 and M33. This feature
is characterized by peak column densities (at 48\arcmin\ resolution) of
only log(N$_{HI})~\sim$~17.5. We have been able to confirm this
exceptionally faint emission with a pointed 30 minute observation with
the GBT on 14 June 2003 directed at
$(\alpha,\delta)$~=~(01:20:29,+37:33:33). Despite the much smaller GBT 
beamsize (of only 9 arcmin) the same low column, log(N$_{HI}$)~=~17.5,
was detected at high significance toward this apparent peak in our
survey data, suggesting a very diffuse gas distribution.  

We have isolated the kinematically continuous features near M31 and M33 in
Fig.~\ref{fig:m31m33}. A handful of faint discrete features to the north-east
of M31 are also included in this figure. These features have velocities
consistent with an M31 population (see Fig.~\ref{fig:vall}) but are not
kinematically continuous with the diffuse component. Similar discrete features
to the south-west of M31 have been excluded because of the greater chance of
confusion with the Magellanic Stream. As noted previously, the bridge can not
be traced as far south as M33, due to foreground confusion from the
Galaxy. For the same reason, the bright disk of M31 is also partially
truncated toward the north-east, giving it a lop-sided appearance relative to
the bridge in Fig.~\ref{fig:m31m33}. Foreground confusion does not hamper
detection of the anti-M33 continuation of the bridge to the north-west of M31
since the velocity gradient of the bridge is toward more negative LGSR (and
GSR) velocity. The total projected extent of this filamentary feature is
about 20\deg, corresponding to 260~kpc. 

Filamentary components extending between massive galaxies are a conspicuous
prediction of high resolution numerical models of structure formation
(eg. Dav\'e et al. \cite{dave99}, \cite{dave01}). Such calculations suggest
that in the current epoch, cosmic baryons are almost equally distributed by
mass amongst three components: (1) galactic concentrations, (2) a warm-hot
intergalactic medium (WHIM) and (3) a diffuse intergalactic medium. These
three components are each coupled to a decreasing range of baryonic
over-density: $log(\rho_{\sc H}/\overline \rho_{\sc H})>3.5$, 1--3.5, and $<$ 1
and are probed by QSO absorption lines with specific ranges of neutral column
density: $log(N_{HI})~>~18$, 14--18, and $<$ 14. The neutral fraction is
thought to decrease with decreasing column density from about 1\% at
log(N$_{HI}$)~=~17, to less than 0.1\% at log(N$_{HI}$)~=~13. While a wide
range of physical conditions can be found within galaxies, the WHIM is thought
to be a condensed shock-heated phase with temperature in the range
10$^5$--10$^7$~K, while the diffuse IGM is predominantly photo-ionized with
temperature near 10$^4$~K. It seems quite conceivable that the M31/M33
filament is the neutral manifestation of such a ``cosmic web'' which
contributes to the ongoing fueling of both galaxies. If we assume that the
overdensity being traced by this structure is $log(\rho_{\sc H}/\overline
\rho_{\sc H}) \sim 3$ (cf. Fig.~10 of Dav\'e et al. \cite{dave99}) then we can
calculate the associated baryonic mass from,
\begin{equation}
M_{Bar} = \Omega_{Bar} {\rho_{\sc H} \over { \overline \rho_{\sc H}} }
{ {3 H_0^2} \over { 8 \pi G} } V
\label{eqn:msim}
\end{equation}
where V is the volume of the structure. Taking $\Omega_{Bar}=0.03$, $H_0=65$
\kms-Mpc$^{-1}$ and $V~=~260\times26\times26$~kpc yields
$M_{Bar}~\sim~6\times10^8$~M$_\odot$ and presumably
$M_{DM}~\sim~6\times10^{9}$~M$_\odot$.  While the apparent baryonic reservoir
in such a scenario is quite substantial, the associated dark mass is less than
1\% of that associated with M31.

A second scenario for the origin of this feature might be a major tidal
interaction between M31 and M33 at some time in the past. If we adopt
distances, D$_{M31}$~=~790 and D$_{M33}$~=~840~kpc (Freedman et
al. \cite{free01}), then the physical separation of the two systems is 218~kpc
and the angle between a radial vector connecting M33 to the Sun with the
vector connecting M33 and M31 is about 70\deg. If we assume, as seems
plausible from test particle simulations in the Local Group potential
(eg. Blitz et al. \cite{blit99}), that the true space velocity of M33 is
directed in the (anti-)M31 direction, then we can estimate it's sign and
magnitude from the observed radial velocity along the M33 to Sun vector. The
measured systemic velocities of M31 and M33 in various reference frames are
summarized in Table~\ref{tab:prev}. The GSR velocity of M33,
V$_{GSR}$~=~$-$44~\kms, is the relevant one, given our heliocentric
perspective. It seems likely that M33 is {\it approaching\ } M31 with a
velocity of about 44/$cos(70\deg)$~=~130~\kms. This would seem to argue
against even a moderately recent (within 2~Gyr) interaction for which a {\it
receding\ } velocity might have been an indication.

\subsection{The M31 HVC Populations}

A major step forward in delineating the M31 HVC populations was made in
our GBT survey (Thilker et al \cite{thil03}). Evidence was found for
three distinct components of high velocity \hi\ within the
95$\times$95~kpc field imaged with the GBT: (1) Several features were
identified of likely tidal origin, including a filamentary component
having partial spatial correspondence with the ``giant'' stellar stream
(Ibata et al \cite{ibat01}, Ferguson et al. \cite{ferg02}, McConachie
et al. \cite{mcco03}). (2) A ``halo'' component of diffuse filamentary
features was detected concentrated at the M31 systemic velocity. (3) A
population of about 20 discrete HVC features was detected within the
7$\times$7\deg\ survey region. The discrete HVC's roughly follow the
kinematic pattern of outer disk rotation and display a correlation of
apparent \hi\ mass with internal line-width, consistent with a dark- to
neutral gas-mass ratio of about 100:1. 

All three of these HVC components are apparent in our wide-field
survey, albeit with reduced point-source sensitivity and angular
resolution. The much wider field-of-view and higher brightness
sensitivity for diffuse features of the current work has
permitted several important additional conclusions to be drawn
regarding the HVC population of M31: (1) The
extreme velocity range of discrete features (comparable with M31
rotation) is only detected within about 12\deg (corresponding to
160~kpc) of M31, despite a much larger FOV over which such features
could have been seen. (2) The systemic velocity ``halo'' component
appears to extend into a diffuse bridge connecting the systemic
velocities of M31 and M33, while continuing in the anti-M33 direction for
an additional $\sim$150~kpc. (3) The M31 HVC populations appear
to be morphologically and kinematically distinct from the Magellanic
Stream, although there is undoubtedly some confusion of discrete features
which happen to lie toward the ``eastern loop''.
 
\begin{figure}
\resizebox{\hsize}{!}{\includegraphics{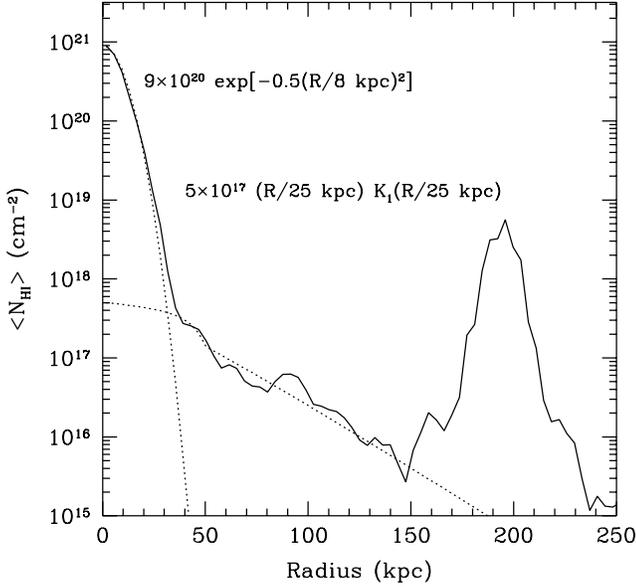}}
 \caption{Azimuthally averaged \hi column density in the vicinity of
 M31 as function of radius. The secondary peak near 200~kpc projected
 distance is due to M33. A Gaussian of 8~kpc dispersion and peak
 log(N$_{HI}$)~=20.95 provides an adequate description of the central
 disk. The extended circum-galactic component can be well-fit with a
 modified Bessel function of 25~kpc scale-length and peak
 log(N$_{HI}$)~=17.7 as discussed in the text.}
\label{fig:nhrad}
\end{figure}

\begin{figure}
\resizebox{\hsize}{!}{\includegraphics{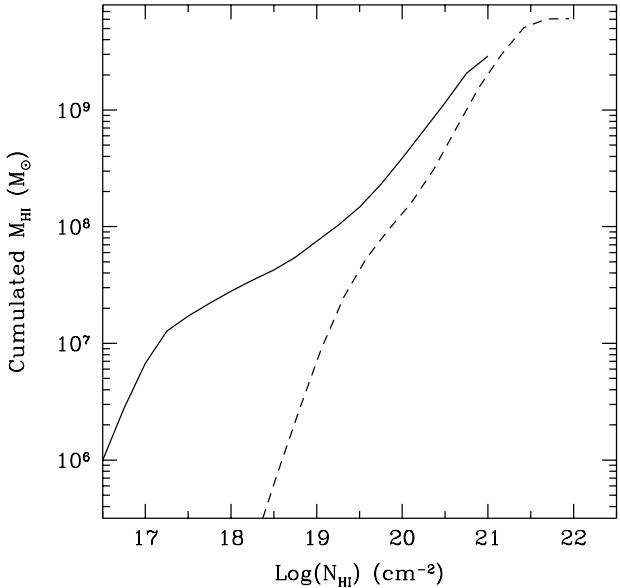}}
 \caption{Cumulated \hi mass in the vicinty of M31 below the indicated
   column density. The solid line is taken from the current wide-field
   survey for the features depicted in Fig.~\ref{fig:m31m33}. The
   dashed line is taken from our high resolution mosaic (Braun et
   al. \cite{brau02}) of the inner 80 by 40~kpc. }
\label{fig:mofnhi}
\end{figure}

It is instructive to consider what the expected properties might be of
a circum-galactic population of low-mass dark-matter halos. Sternberg
et al. (\cite{ster02}) have calculated the hydrostatic and ionization
structures for such objects under a variety of assumptions about the
confining medium. Bound neutral cores can only be realized with dark
matter masses larger than about M$_{\rm DM}>$10$^8$M$_\odot$,
corresponding to maximum rotation velocities, $v_{\rm max}>$10
\kms. This is because a shielding column of warm ionized gas (WIM, with
temperature 10$^4$K) must be retained, before a warm neutral component
(WNM, with 8000$<$T$<$10$^4$K) can survive. Under appropriate
conditions, a cool neutral core (CNM, with 50$<$T$<$200~K) can condense
within this nested structure. The \hi\ mass associated with these
low mass objects depends sensitively on the external pressure,
P$_{\rm HIM}/k$, of the hot ionized gas (HIM, with
10$^{5.5}<$T$<$10$^{6.5}$K) within which they are embedded. For M$_{\rm
DM}=10^8$M$_\odot$, M$_{\rm HI}$ varies approximately quadratically
from about $10^3$ to $10^7$ M$_\odot$ with the external pressure,
P$_{\rm HIM}/k=$1 to 100 cm$^{-3}$K.  For a realistic external
pressure of a few 10's of cm$^{-3}$K, the minimum expected \hi\ mass
might be about 10$^5$M$_\odot$. Given the steeply rising mass function
at low M$_{\rm DM}$, N(M$_{\rm DM}$)$\propto$M$_{\rm DM}^{-2}$, this
will also be the most commonly occurring \hi\ mass. In their eqn.~38,
Sternberg et al. give the expected number of sub-halos within a
distance, $d$, of the parent halo of mass, $M_{vir,p}$, which exceed a
given $v_{\rm max}$, based on the numerical simulations of Klypin et
al. (\cite{klyp99}),
$$N(>v_{\rm
  max},<d)=1.06\times10^3\left({{M_{vir,p}}\over{10^{12}M_\odot}}\right)$$ 
\begin{equation}
\hfill
\times\left({{v_{\rm max}}\over{\rm
    8~km~s^{-1}}}\right)^{-2.75}\left({{d}\over{\rm 1~Mpc}}\right).
\label{eqn:nsim}
\end{equation}
Thilker et al. (\cite{thil03}) have already pointed out that the 20
discrete objects with \hi\ masses in the range $10^5$ and $10^7$
M$_\odot$ detected in the GBT survey field are in moderately good
agreement with the expected number, $N=25$, predicted by the above
equation (with $v_{\rm max}>10$ \kms, $M_{vir,p}=10^{12}$ M$_\odot$
and $d=40$~kpc).  The 8$\sigma$ completeness limit of the GBT survey
(for a 15\arcmin, 32~\kms\ source extent) is 2$\times10^5$M$_\odot$ at
the M31 distance, suggesting that the GBT detections are already quite
incomplete at the low mass end. The 8$\sigma$ completeness limit of the
WSRT wide-field survey (for a spatially unresolved, 32~\kms\ source
extent) is 4.5$\times10^5$M$_\odot$. The relative completeness of the
current and the GBT surveys is confirmed within the 7$\times$7\deg
region of overlapping coverage. Only the brighter (and unconfused) GBT
detections are apparent in our wide-field survey. 

Extending the survey coverage from $d=40$~kpc, probed by our GBT data,
to $d=160$~kpc in our wide-field data is predicted (by
eqn.~\ref{eqn:nsim}) to increase the number of such objects by a factor
of four to about 100.  However, the implication of our moderately high
completeness limit on large scales, M$_{\rm
HI}$=4.5$\times10^5$M$_\odot$, is that such a population of low mass
discrete components will require higher sensitivity for their
detection. This would be exacerbated in practise by a likely radial
gradient in P$_{\rm HIM}$ in the extended halo of M31. A decline in
P$_{\rm HIM}$ at larger radii would result in a rapid decline in the
associated \hi\ mass of any dark matter sub-halo.

Although our sampling of the HVC population of M31 is known to be incomplete,
due to the effects of confusion with both the Galaxy and the Magellanic
Stream, as well as our limited sensitivity, we can begin to assess the
properties of the combined large-scale distribution. In Fig.~\ref{fig:nhrad}
we plot the azimuthally averaged column density of currently detected features
as a function of projected distance to M31. Only the subset of features
apparently associated with M31 and M33 (and illustrated in
Fig.~\ref{fig:m31m33}) have been used in making this figure. The bright
central disk component can be characterized by a Gaussian of 8~kpc dispersion
and peak column density log(N$_{HI}$)~=20.95. The secondary peak in the figure
near 200~kpc radius is due to the bright disk emission of M33. The
distribution of circum-galactic gas can be well-described by the projection of
a radial exponential of the neutral gas volume density, $n_{\rm HI}$, with a
radial scale-length, $h$~=~25~kpc and a peak log(N$_{HI}$)~=17.7.  Taking
\begin{equation}
n_{\rm HI}(r) = n_{\rm o}e^{-r/h}
\label{eqn:no}
\end{equation}
in terms of the radial distance, $r$; the corresponding projected
distribution of \hi column density,
\begin{equation}
N_{\rm HI}(r) = 2 h n_{\rm o} {r \over h}K_1 \biggl( { r \over
h}\biggr),
\label{eqn:No}
\end{equation}
where $K_1$ is the modified Bessel function of order 1. The peak column
density of the distribution is $N_{\rm HI}(0) = 2 h n_{\rm o}$ and the
integrated mass is given by,
\begin{equation}
M_{\rm HI} = 24 \pi n_{\rm o} m_{\rm HI} h^3 = 12 \pi m_{\rm HI}
N_{\rm HI}(0) h^2
\label{eqn:mint}
\end{equation}
The projected-exponential (of eqn.~\ref{eqn:No}) provides a better fit
to the observed profile than a simple exponential or Gaussian,
particularly for the ``shoulder'' feature near 50~kpc, where much of
the mass of the distribution is concentrated. This form has the
additional advantage (over a 1-D exponential, for example) of lending
itself to a straightforward physical interpretation. The integrated
\hi\ mass of the disk Gaussian component is 2.9$\times10^9$M$_\odot$
and of the circum-galactic exponential is 9.4$\times10^7$M$_\odot$. The
disk component falls about 40\% short of the total \hi\ mass of M31 as
discussed further below.

Although the circum-galactic component of M31 is spatially quite extended, it
is more centrally concentrated than the best-fitting models for a Galactic
sub-halo population of De Heij et al. (\cite{dehe02c}), which called for a
Gaussian spatial dispersion between about 150-200~kpc. Those model fits were
constrained by the observed properties of the all-sky CHVC samples of Putman
et al. (\cite{putm02}) and De Heij et al. (\cite{dehe02a}) which appear to be
completely dominated by the sub-population associated with the Galaxy. In
predicting the attributes of a corresponding M31 population, De Heij et
al. assumed that identical spatial scale-lengths might apply to both the
Galaxy and M31, and that the total number of M31 sub-halos would be about
twice that associated with the Galaxy, given a factor two higher assumed
virial mass (cf. eqn~\ref{eqn:nsim}). It is conceivable that the
circum-galactic populations of the Galaxy and M31 are actually quite different
and also that the distributions are not spherically symmetric, but rather
more filamentary in character as is quite apparent from
Fig.~\ref{fig:m31m33}.

In Fig.~\ref{fig:mofnhi} we plot the cumulated \hi\ mass due to
lines-of-sight with less than the indicated column density. The
distribution depicted in Fig.~\ref{fig:m31m33} leads to the solid line
in the figure. For comparison we also plot the cumulated mass detected
in our complimentary high resolution survey (Braun et
al. \cite{brau02}) of the central 80$\times$40~kpc (major$\times$minor
axis) area as a dashed line. The apparent rapid roll-off in the curves
below about log(N$_{HI}$)~=~17.2 (for the wide-field survey) and
log(N$_{HI}$)~=~19.2 (for the high-resolution survey) is an artifact of
the limited sensitivity. The solid curve only reaches about 60\% of the
total mass given by the dashed curve, since the integration in velocity
has been truncated where foreground confusion sets in for the
wide-field data. This truncation could be circumvented in the
high-resolution data by careful spatial filtering. The excess \hi\ mass
due to low column density emission in the extended M31 environment (out
to 160~kpc) amounts to about $10^8$ M$_\odot$. This is in good
agreement with the integral of the projected exponential function noted
above. Relative to a total M$_{HI}$ of about $5\times10^9$ M$_\odot$,
this is only about 2\%.

\begin{figure}
\resizebox{\hsize}{!}{\includegraphics{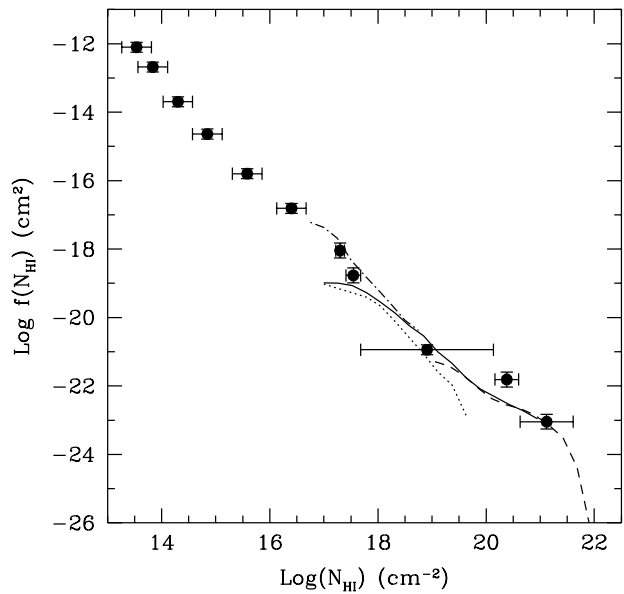}}
 \caption{The distribution function of \hi column density
 due to M31 and it's environment. The data from three \hi surveys of
 M31 are combined in this figure to probe column densities over a
 total range of some five orders of magnitude. The dashed line is from
 the WSRT mosaic (Braun et al. \cite{brau03b}) with 1$^\prime$
 resolution over 80$\times$40~kpc, the dotted and solid lines from our
 GBT survey (Thilker et al. \cite{thil03}) with 9$^\prime$ resolution
 over 95$\times$95~kpc and the dot-dash line from the current work with
 48$^\prime$ resolution out to 150~kpc radius. The filled circles with
 errorbars are the low red-shift QSO absorption line data as tabulated
 by Corbellli \& Bandiera (\cite{corb02}). }
\label{fig:fnhi}
\end{figure}

\subsection{Circum-galactic \hi\ and QSO Absorption Lines}

In addition to the current wide-field survey of the M31 environment
(800$\times$400~kpc at 11~kpc resolution), we have imaged the central
95$\times$95~kpc with the GBT (Thilker et al. \cite{thil03}) at 2~kpc
resolution and the central 80$\times$40~kpc (major$\times$minor axis)
area at up to 50~pc resolution in our WSRT synthesis mosaic (Braun et
al. \cite{brau02}, \cite{brau03b}). The combined database of \hi\ detections
associated with M31 enables compilation of the distribution function of
\hi\ column density spanning an unprecedented range of almost five
orders of magnitude from log(N$_{HI}$)~=~17.2 to log(N$_{HI}$)~=~21.9.
We plot the combined distribution function in Fig.~\ref{fig:fnhi},
where the dashed line is drawn from our synthesis mosaic at 1\arcmin\
resolution, the dotted and solid lines from our GBT survey and the
dot-dash line from the current work. The distinction between the dotted
and solid curves for the GBT data is that all features are included in
the solid line plot, while only the peculiar velocity features that
could be clearly isolated from normal disk emission are indicated by
the dotted line. Each curve displays an apparent turn-over at low
column densities which reflects the finite column density sensitivity
of the survey. For our synthesis mosaic data this sets in below
log(N$_{HI}$)~=~19.3, for the GBT survey below about log(N$_{HI}$)~=~18
and for the wide-field survey below about log(N$_{HI}$)~=~17. The
milder flattening of the GBT data below log(N$_{HI}$)~=~18.2 is very
likely due to an insufficient field-of-view to fully sample the
distribution.

The curves in  Fig.~\ref{fig:fnhi} were calculated from,
\begin{equation}
f(N_{\rm HI}) = {c \over H_0}{\rho_{av} \over \rho_{M31}} \theta_{M31}
{A(N_{\rm HI}) \over dN_{\rm HI}}\ \ \  {\rm cm}^2
\label{eqn:fnhi}
\end{equation}
where 
\begin{equation}
\theta_{M31} = \theta_* {\rm ln}(10) ({\rm M_{M31}}/ {\rm
M_*})^{\alpha+1} {\rm exp}(-{\rm M_{M31}/M_*})
\label{eqn:tm31}
\end{equation}
and $\rho_{M31} = \theta_{M31} {\rm M_{M31}}$. $A(N_{\rm HI})$ is the
surface area subtended by \hi\ in the column density interval $dN_{\rm
HI}$ centered on $N_{\rm HI}$. The HIMF parameters were taken from
Zwaan et al. (\cite{zwaa03}); namely a faint-end slope, $\alpha=-1.30$,
characteristic \hi\ mass, log(M$_*/$M$_\odot)=9.79$ and normalization,
$\theta_*=8.6\times10^{-3}$ Mpc$^{-3}$, where $H_0=75$
km$^{-1}$Mpc$^{-1}$ has been assumed throughout. The term, $\rho_{av} /
\rho_{M31}$ in eqn.~\ref{eqn:fnhi} is introduced to normalize our
measurements for a single galaxy, M31, to the global average density of
\hi\ in galaxies, $\rho_{av}= 6.1\times10^7h_{75}$M$_\odot$Mpc$^{-3}$,
also taken from Zwaan et al. Since the \hi\ mass of M31 places it close
to M$_*$, this is a relatively small upward correction of only 30\%.

Our distribution function can be compared with previous \hi\ work by
Ryan-Weber et al. (\cite{ryan03}) and Rao \& Briggs (\cite{rao93}) who
considered \hi\ selected samples of galaxies. While these studies are
superior in their sampling of a range in galaxy type and \hi\ mass,
they are limited by sensitivity to about log(N$_{HI})>19.5$ and by
angular resolution to log(N$_{HI})<21$. Of course, for log(N$_{HI})>21$
there is a high liklihood of significant optical depth in the \hi\
emission line (Braun \& Walterbos \cite{brau92}, Braun \cite{brau97}),
so that even our high resolution estimates must be regarded as lower
limits in this column density range.

We also compare our distribution function to the low red-shift QSO absorption
line data, as tabulated by Corbelli \& Bandiera (\cite{corb02}), in
Fig.~\ref{fig:fnhi}. The filled circles with error bars show the QSO data
together with the column density ranges over which they were defined. Four of
the five measurement points show very good agreement with our distribution
function. Only the QSO measurement near log(N$_{HI})=20.4$ suggests an excess
probability of absorption by more than a factor of four relative to that seen
toward M31. The \hi\ distribution function of Ryan-Weber et
al. (\cite{ryan03}) is in good agreement with ours in this column density
range, making it unlikely that this apparent deficit is simply a peculiarity
of the \hi\ distribution of M31. An explanation for this apparent discrepancy
may lie in the misclassification of some absorption line systems (eg. Turnshek
\& Rao \cite{turn02}).  In Fig.~\ref{fig:fnhi} we plot the low red-shift
optical distribution function over it's entire measured range, to put our
result into a broader perspective. If the distribution we detect around M31 is
at all representative, then it can account very well for the QSO absorption
line statistics.

The QSO absorption line data has a profound implication for studies of the
\hi\ universe. By going down in column density sensitivity from
log(N$_{HI})=19$, such as available in the deepest current imaging studies of
nearby galaxies, to log(N$_{HI})=17$, the surface covering factor of \hi\
emission features should increase by about a factor of 30. The QSO absorption
line statistics allow this prediction to be made with great confidence. There
has been debate about how this gas might be distributed, but it seemed
intimately linked in some way to L$_*$ (and therefore M$_*$) galaxies,
possibly in the form of a cloud of low-mass satellites or an extension of the
L$_*$ galaxy disk. In at least one instance we have been able to confirm this
predicted increase in surface covering factor at low N$_{HI}$ and demonstrate
in an image how the system is related to the host galaxy. As noted above,
the detected morphology is suggestive of the ``cosmic web'' of
filaments which are predicted in recent numerical simulations of structure
formation (eg. Dav\'e et al. \cite{dave99}). 

\section{Summary}
\label{sec:summary}

Our wide-field \hi\ survey of 1800 deg$^2$ around M31 has provided a
number of insights into the low N$_{HI}$ sky. Peculiar velocity neutral
gas, in excess of our 3$\sigma$ limit of
N$_{HI}=1.5\times10^{17}$cm$^{-2}$, is seen from 29\% of our survey area.
This is comparable to the 37\% detection rate of Lockman et
al. (\cite{lock02}) for high velocity gas, with
N$_{HI}>3\times10^{17}$cm$^{-2}$, toward 860 semi-random directions at
$\delta>-44^\circ$. 

Our improved sensitivity over previous surveys has permitted detection of an
order of magnitude more discrete HVC features in this region, increasing their
number from 9 to 95. The vast majority of these discrete features (83\%) are
isolated in position and velocity at a level of
N$_{HI}=1.5\times10^{17}$cm$^{-2}$ from any adjoining HVC complexes, and may
thus be termed CHVCs. While the distributions of angular size and velocity
width of these features are comparable to what has been seen previously, 
the distribution of flux density appears to have an inflection point from a
steep power-law index ($-2.15$ in linear units) above 12~Jy-\kms\ to a much
flatter distribution down to our $8\sigma$ completeness limit of 6~Jy-\kms\
(see Fig.~\ref{fig:hvc}). Such an inflection point may be an indication that
the spatial distribution of discrete HVCs is not isotropic. Sternberg et
al. (\cite{ster02}) have already pointed out that the total number of
cataloged CHVCs (in the HIPASS and LDS samples) is consistent with that
expected from the numerical $\Lambda$CDM simulations and parameterized in
eqn.~\ref{eqn:nsim} if the population has a characteristic distance of about
150~kpc. At such a characteristic distance, 12~Jy-\kms\ corresponds to about
M$_{HI}=6\times10^4$M$_\odot$. 

A prominent diffuse feature detected in our survey field is a northern
extension of the Magellanic Stream, as seen in
Fig.~\ref{fig:magstm}. At northern declinations the MS is bifurcated
into a linear feature which we term the ``western filament'' as well as
an arc-like structure we term the ``eastern loop''. Typical column
densities in these features are about $3\times10^{17}$cm$^{-2}$, while
the line-of-sight velocity is essentially single-valued and varies
smoothly with position as shown in Fig.~\ref{fig:vall}.  Both the
location and kinematics of these extensions agree well with the
tidally-dominated simulations of Gardiner \& Noguchi (\cite{gard96})
and Gardiner (\cite{gard99}). In these models the ``eastern loop''
corresponds to the apo-galacticon portion of the LMC/SMC orbit, near
125~kpc, where the clouds were located some 0.9~Gyr previously.

Adjacent to the MS is the HVC feature known as Wright's Cloud. We
detect faint extensions of this complex reaching about 6\deg\ to the
south-east and high-light the peculiar internal kinematics of the
bright core region. Two principal kinematic axes are present which are
essentially perpendicular. Velocity gradients of at least 65 and
45~\kms\ are present along these overlapping axes. We suggest that a
plausible explanation of this feature may involve an \hi\ condensation
of about 10$^7$M$_\odot$ on a dark sub-structure of 10$^9$M$_\odot$ in
the extended galactic halo at a distance of about 120~kpc. Higher
resolution imaging should help to better constrain the kinematics and
likely origin of this source.

In the immediate vicinity of M31 we detect a system of very faint
discrete \hi\ features with radial velocities that span most of the
range corresponding to rotation in the M31 disk. Such features are only
detected out to a radius of about 12\deg, corresponding to 160~kpc from
M31 with our current sensitivity (an 8$\sigma$ completeness limit of
$4.5\times 10^5$M$_\odot$). The inner $95\times95$~kpc of this
distribution has been imaged with higher resolution and point source
sensitivity in our complimentary survey with the GBT (Thilker et
al. \cite{thil03}) (for which the 8$\sigma$ completeness limit is
$2\times 10^5$M$_\odot$). With the higher resolution of the GBT
(9\arcmin\ corresponding to 2~kpc) it is clear that a subset of these
discrete features are tidal in origin, while others do not appear to
be. Only one feature in this class (Davies' Cloud) has an apparent
\hi\ mass as high as 10$^7$M$_\odot$, with the remainder in the range
$10^5\rightarrow10^6$M$_\odot$. This discrete population appears to be
an excellent candidate for the \hi\ associated with low mass dark
matter satellites of M31. The detected numbers of objects and their
associated \hi\ mass are in good agreement with expectations (Sternberg
et al. \cite{ster02}).

A diffuse filamentary distribution of \hi\ is detected at the systemic
velocity of M31. The filament extends in the direction of M33 (at a projected
distance of 200~kpc), varying in line-of-sight velocity toward the M33
systemic velocity. The filament is also detected out to about 150~kpc in the
anti-M33 direction. Peak column densities in this feature amount to only about
$3\times10^{17}$cm$^{-2}$ in an 11~kpc beam. A similarly low column density
has been detected in a deep pointed observation of a local peak with the much
smaller GBT beam (2~kpc), suggesting that the filament is indeed extremely
diffuse. This feature appears to be an extension of the ``halo'' component
concentrated at the M31 systemic velocity that was seen in our GBT survey
(Thilker et al. \cite{thil03}). This structure is suggestive of \hi\
associated with a filament of coronal gas, either in the form of localized
neutral condensations are simply as a trace neutral constituent of gas with a
high degree of ionization. If we are detecting \hi\ from a warm-hot
inter-galactic medium (WHIM) filament then it is likely to contain some
$6\times10^8$~M$_\odot$ of baryons and presumably $6\times10^{9}$~M$_\odot$ of
dark matter.

The distribution of \hi\ in the extended environment of M31 can be
described by a projected exponential function ($x K_1(x)$) of 25~kpc
scale-length and $5\times10^{17}$cm$^{-2}$ peak column density. The
combined distribution function of \hi in the M31 disk and environment
has been constructed over the column density range log(N$_{HI}$)~=~17.2
to log(N$_{HI}$)~=~21.9 by utilizing all three of our complimentary
surveys spanning resolutions of 50~pc to 11~kpc. The composite
distribution function provides very good agreement with the low
red-shift QSO absorption line data (as tabulated by Corbelli \&
Bandieri \cite{corb02}) over this entire range, with the 
exception of the QSO point near log(N$_{HI})=20.4$, for which the M31
data fall short by about a factor of four. 

Our survey has demonstrated that the predicted increase in surface
covering factor of \hi\ at low column density (by more than three
orders of magnitude) implied by the QSO absorption line statistics is
borne out in practise. Extremely diffuse filaments of the ``cosmic
web'' in the environment of $M_*$ galaxies are open to direct imaging
with sensitive observations of the $\lambda$21cm line.

\begin{table*}
\renewcommand{\thefootnote}{\thempfootnote}
\caption{ Properties of detected high velocity clouds
}
\label{tab:hvcs}
{\setlength{\tabcolsep}{4pt}
\begin{tabular}{rrccrrrrrrrl}

\hline
{\#}& {Name}                     & {RA$_{2000}$}
                         & {Dec$_{2000}$}
                         & {V$_{\rm HEL}$}
                         & {FWHM}
                         & {N$_{HI}$}
                         & {MAJ}
                         & {MIN}
                         & {PA}
                         & {S$_{int}$\ }
                         & {Catalog IDs$^{\mathrm{a}}$}\\
{\ }&{\sc ddddd lll.l$-$bb.b$-$vvv}                       & {$\rm(h\ m\ s)$}
                         & {$\rm(^\circ\ ^\prime\ ^{\prime\prime})$}
                         & \multispan2{\ \ (\kms)}
                         & {(10$^{18}$cm$^{-2}$)} 
                         & {(\deg)}
                         & {(\deg)}
                         & {(\deg)}
                         & {(Jy-\kms)} 
                         & {\ } \\
(1)&(2)&(3)&(4)&(5)&(6)&(7)&(8)&(9)&(10)&(11)&(12)\\
\hline
 1&CHVC 086.6$-$28.7$-$392&22 32 26&24 14 04&$-$400& 48.0&0.90&2.17&1.47&175& 33.6&                           \\
 2&CHVC 089.7$-$33.1$-$215&22 52 36&22 00 20&$-$221& 33.4&0.73&1.64&0.93&139& 13.0&            	       \\
 3&CHVC 091.5$-$26.6$-$395&22 43 24&28 20 52&$-$402& 25.8&0.47&0.88&0.78&166&  3.8&            	       \\
 4&CHVC 091.7$-$35.7$-$400&23 04 31&20 37 04&$-$405& 40.9&5.89&1.14&0.96&  8& 75.4&{\sc dbb326,ww493}	       \\
 5&CHVC 092.5$-$23.6$-$411&22 39 32&31 22 32&$-$419& 35.1&0.50&1.79&0.78&151&  8.2&            	       \\
 6&CHVC 092.6$-$30.7$-$404&22 56 48&25 17 48&$-$410&103.0&4.14&2.15&2.12&  3&220.9&            	       \\
 7&CHVC 093.4$-$11.9$-$485&22 10 50&41 31 57&$-$495& 36.1&4.26&0.94&0.86& 99& 40.3&            	       \\
 8&CHVC 093.8$-$09.3$-$342&22 04 12&43 53 12&$-$352& 32.6&0.57&0.96&0.78&160&  5.0&            	       \\
 9&CHVC 094.8$-$11.7$-$447&22 16 09&42 32 49&$-$457& 74.7&0.70&1.56&0.78&111& 10.0&            	       \\
10&CHVC 095.1$-$17.0$-$394&22 33 02&38 15 33&$-$403& 34.6&0.45&1.01&0.78&102&  4.1&            	       \\
11&CHVC 095.1$-$17.9$-$286&22 35 12&37 26 42&$-$294& 33.8&0.36&1.02&0.78&  8&  3.4&            	       \\
12& HVC 095.1$-$06.1$-$370&21 59 16&47 14 00&$-$380& 31.8&1.60&1.09&0.78& 94& 15.9&            	       \\
13&CHVC 095.5$-$21.9$-$370&22 46 47&34 12 04&$-$378& 40.6&0.64&0.86&0.78& 23&  5.0&            	       \\
14&CHVC 095.6$-$14.1$-$470&22 26 37&40 57 35&$-$479& 29.4&1.16&0.84&0.78& 90&  8.9&            	       \\
15& HVC 096.4$-$99.5$-$354&22 16 33&45 12 55&$-$364& 29.0&2.51&1.22&0.83&142& 29.7&            	       \\
16&CHVC 096.7$-$12.9$-$454&22 28 10&42 30 14&$-$463& 45.5&1.13&0.93&0.78&110&  9.6&            	       \\
17&CHVC 096.9$-$16.0$-$316&22 37 41&40 00 10&$-$324& 34.8&0.57&1.77&0.85&149& 10.0&            	       \\
18&CHVC 098.4$-$13.7$-$449&22 38 20&42 41 41&$-$458& 32.6&0.73&0.92&0.78&114&  6.1&            	       \\
19&CHVC 098.1$-$16.8$-$345&22 45 07&39 54 56&$-$353& 39.5&0.59&1.17&0.78&106&  6.3&            	       \\
20&CHVC 098.8$-$31.1$-$420&23 18 25&27 24 32&$-$425& 26.3&0.44&1.91&0.78&  8&  7.7&            	       \\
21&CHVC 100.0$-$36.8$-$375&23 32 25&22 29 33&$-$379& 32.0&1.24&1.89&1.44& 10& 39.5&            	       \\
22&CHVC 100.7$-$13.8$-$375&22 48 57&43 45 13&$-$383& 45.6&1.90&1.13&0.78&121& 19.6&            	       \\
23&CHVC 101.3$-$22.6$-$439&23 11 40&36 04 14&$-$446& 33.7&0.74&0.96&0.78&138&  6.5&            	       \\
24&CHVC 102.6$-$28.6$-$427&23 28 07&30 56 40&$-$432& 34.8&0.71&1.62&1.11& 58& 14.9&            	       \\
25&CHVC 103.0$-$32.8$-$436&23 36 08&27 07 45&$-$440& 26.3&0.74&1.38&0.85&105& 10.2&            	       \\
26&CHVC 104.0$-$23.1$-$446&23 24 08&36 30 09&$-$452& 34.1&1.56&1.31&0.89&118& 21.3&            	       \\
27&CHVC 104.8$-$18.5$-$455&23 19 18&41 07 21&$-$462& 29.7&1.65&0.93&0.92&166& 16.5&            	       \\
28& HVC 104.8$-$38.4$-$402&23 50 18&22 17 55&$-$405& 28.0&1.44&0.94&0.81& 11& 12.8&            	       \\
29&CHVC 105.0$-$24.6$-$430&23 31 05&35 25 51&$-$436& 24.5&0.50&2.50&0.85&151& 12.4&            	       \\
30&CHVC 105.4$-$28.4$-$413&23 38 31&31 59 42&$-$418& 38.9&0.55&0.84&0.78& 90&  4.2&            	       \\
31&CHVC 105.9$-$34.0$-$430&23 48 17&26 48 06&$-$434& 35.4&1.45&1.23&1.07& 14& 22.3&            	       \\
32&CHVC 106.4$-$37.3$-$413&23 54 06&23 43 28&$-$416& 29.0&1.25&0.92&0.78&109& 10.5&            	       \\
33&CHVC 107.5$-$29.8$-$422&23 48 44&31 10 45&$-$426& 33.1&7.42&1.30&1.02& 65&115.2&{\sc dbb385,ww437,bb22}    \\
34&CHVC 107.6$-$33.9$-$379&23 54 17&27 15 11&$-$382& 69.6&0.91&0.85&0.82&140&  7.4&            	       \\
35&CHVC 108.2$-$21.4$-$398&23 39 55&39 26 43&$-$403& 28.6&6.13&1.05&0.82&127& 61.8&{\sc dbb387,ww389,bb23}    \\
36&CHVC 109.0$-$35.0$-$430&00 00 28&26 31 39&$-$433& 30.5&0.47&1.30&0.84&  7&  6.0&            	       \\
37&CHVC 109.1$-$31.8$-$330&23 57 24&29 36 46&$-$333& 23.4&0.48&1.56&1.18&138& 10.3&            	       \\
38&CHVC 109.2$-$19.8$-$433&23 42 09&41 11 49&$-$439& 35.3&3.30&0.88&0.84& 65& 28.6&            	       \\
39&CHVC 109.5$-$30.6$-$389&23 57 29&30 48 57&$-$393& 27.2&0.48&1.15&0.74& 76&  4.8&            	       \\
40&CHVC 110.1$-$32.3$-$365&00 01 33&29 20 14&$-$368& 31.9&1.67&1.23&0.94& 36& 22.6&            	       \\
41& HVC 110.5$-$24.4$-$407&23 54 44&37 05 02&$-$411& 26.8&0.96&2.13&1.16& 72& 27.8&            	       \\
42&CHVC:110.8$-$16.4$-$418&23 45 25&44 53 58&$-$424& 34.8&2.93&1.82&0.99& 35& 61.8&             	       \\
43&CHVC 110.9$-$29.6$-$178&00 02 13&32 07 01&$-$181& 33.3&3.19&1.82&1.04& 15& 70.7&            	       \\
44&CHVC 111.2$-$29.3$-$366&00 03 00&32 30 22&$-$369& 30.5&0.49&1.44&0.96& 12&  7.9&            	       \\
45& HVC 111.6$-$23.8$-$413&23 58 35&37 51 52&$-$417& 31.8&2.84&2.22&1.01& 30& 74.5&            	       \\
46&CHVC:111.9$-$15.6$-$407&23 49 57&45 57 15&$-$413& 36.0&1.56&1.75&0.94& 63& 30.0&            	       \\
47&CHVC:112.0$-$14.0$-$371&23 48 19&47 30 10&$-$377& 36.3&0.94&1.24&0.80& 28& 10.9&            	       \\
48&CHVC:112.6$-$12.9$-$365&23 49 46&48 41 47&$-$371& 30.1&1.85&1.23&1.00& 37& 26.6&            	       \\
49& HVC 112.2$-$22.5$-$407&23 59 52&39 16 07&$-$411& 29.4&3.94&1.32&1.07& 33& 65.1&            	       \\
50&CHVC 112.7$-$26.3$-$373&00 06 05&35 38 33&$-$377& 30.3&0.86&1.27&0.92&175& 11.8&            	       \\
\hline
\end{tabular} }
\begin{list}{}{}
\item[$^{\mathrm{a}}$] Catalog ID's prefaced with DBB refer to the entries
of Table~1 of De Heij et al. \cite{dehe02a}, those prefaced with BB to Table~1
of Braun \& Burton \cite{brau99} and those with WW to Table~1 of Wakker \& 
van Woerden \cite{wakk91}.
\end{list}
\end{table*}

\begin{table*}
\caption{ Properties of detected high velocity clouds (continued)
}
{\setlength{\tabcolsep}{4pt}
\begin{tabular}{rrccrrrrrrrl}
\hline
{\#}& {Name}                     & {RA$_{2000}$}
                         & {Dec$_{2000}$}
                         & {V$_{\rm HEL}$}
                         & {FWHM}
                         & {N$_{HI}$}
                         & {MAJ}
                         & {MIN}
                         & {PA}
                         & {S$_{int}$\ }
                         & {Catalog IDs$^{\mathrm{a}}$}\\
{\ }&{\sc ddddd lll.l$-$bb.b$-$vvv}                       & {$\rm(h\ m\ s)$}
                         & {$\rm(^\circ\ ^\prime\ ^{\prime\prime})$}
                         & \multispan2{\ \ (\kms)}
                         & {(10$^{18}$cm$^{-2}$)} 
                         & {(\deg)}
                         & {(\deg)}
                         & {(\deg)}
                         & {(Jy-\kms)} 
                         & {\ } \\
(1)&(2)&(3)&(4)&(5)&(6)&(7)&(8)&(9)&(10)&(11)&(12)\\
\hline
51&CHVC 113.1$-$25.0$-$384&00 06 39&37 02 37&$-$388&29.9&0.53&1.78&1.13&129& 12.5&            	       \\
52&CHVC 113.7$-$41.9$-$266&00 22 19&20 26 20&$-$267&47.9&1.12&1.21&0.86& 68& 13.6&                            \\
53&CHVC 114.2$-$23.3$-$302&00 10 11&38 49 16&$-$306&28.6&2.07&1.11&0.84& 79& 22.6&            	        \\
54&CHVC 115.4$-$23.1$-$503&00 15 28&39 15 12&$-$507&20.0&0.20&1.46&0.82& 80&  2.8&            	        \\
55&CHVC 116.3$-$28.9$-$376&00 23 40&33 39 00&$-$379&30.1&0.86&1.39&0.83& 76& 11.6&            	        \\
56& HVC 116.6$-$36.6$-$157&00 28 40&25 58 19&$-$158&40.0&2.78&1.62&0.93& 68& 49.0&            	        \\
57& HVC 117.1$-$32.6$-$387&00 28 45&30 02 53&$-$389&31.2&1.17&0.96&0.91&174& 12.0&            	        \\
58&CHVC 117.7$-$27.4$-$332&00 28 38&35 12 22&$-$334&30.6&0.52&1.18&0.87& 35&  6.2&            	        \\
59& HVC 118.4$-$32.6$-$378&00 33 39&30 08 21&$-$380&33.3&4.41&0.93&0.78&103& 37.4&{\sc dbb415}	        \\
60& HVC 118.7$-$29.4$-$367&00 33 45&33 22 26&$-$369&47.7&0.93&1.81&0.82& 95& 16.2&           		        \\
61& HVC 118.5$-$31.3$-$133&00 33 52&31 22 50&$-$135&27.6&3.33&1.93&0.94& 15& 70.7&{\sc dbb416}   		\\
62&CHVC 119.0$-$19.0$-$370&00 30 41&43 44 55&$-$374&32.7&0.70&1.41&1.08& 18& 12.5&            	        \\
63& HVC 119.2$-$30.9$-$384&00 36 15&31 54 12&$-$386&30.1&6.84&0.97&0.85&150& 66.0&{\sc dbb422,ww444,bb29}	\\
64&CHVC 119.7$-$29.9$-$328&00 38 15&32 53 59&$-$330&43.1&2.17&1.10&0.85&130& 23.7&            	        \\
65& HVC 120.2$-$20.3$-$442&00 37 17&42 31 12&$-$445&27.7&9.26&0.97&0.85&121& 89.4&{\sc dbb425}		\\
66&CHVC 120.4$-$35.3$-$361&00 41 57&27 34 13&$-$362&51.0&1.75&1.24&0.89&103& 22.6&            	        \\
67&CHVC 121.3$-$28.5$-$352&00 44 23&34 19 42&$-$354&31.8&0.93&0.87&0.79& 52&  7.5&            	        \\
68&CHVC 121.6$-$18.6$-$268&00 44 18&44 16 26&$-$271&35.3&0.51&0.84&0.78& 90&  3.9&            	        \\
69&CHVC 122.4$-$15.3$-$338&00 48 11&47 33 04&$-$341&26.0&1.19&1.08&0.90&101& 13.5&            	        \\
70&CHVC 122.4$-$38.9$-$329&00 49 32&23 58 38&$-$329&32.3&2.56&1.48&0.96& 90& 42.6&            	        \\
71&CHVC 122.6$-$32.8$-$382&00 49 59&30 02 56&$-$383&29.0&2.23&1.23&0.88&155& 28.3&            	        \\
72&CHVC 122.5$-$39.7$-$358&00 50 06&23 11 18&$-$358&32.0&1.36&1.48&1.01&169& 23.8&            	        \\
73&CHVC 122.9$-$32.0$-$326&00 51 28&30 51 47&$-$327&35.9&4.24&1.13&0.86&  2& 48.2&{\sc dbb432,ww446,bb30}	\\
74&CHVC 123.1$-$42.0$-$364&00 52 00&20 53 01&$-$363&35.2&1.53&1.02&0.84&117& 15.3&            	        \\
75&CHVC 123.1$-$23.6$-$401&00 52 12&39 15 10&$-$403&30.2&0.39&1.38&1.09&146&  6.9&            	        \\
76&CHVC:123.8$-$38.9$-$318&00 54 33&23 55 01&$-$317&39.1&2.04&1.44&1.03& 75& 35.4&           		        \\
77&CHVC 124.3$-$24.1$-$471&00 57 52&38 46 32&$-$473&34.8&0.53&1.94&0.78& 93&  9.4&            	        \\
78& HVC 126.2$-$31.8$-$408&01 04 18&30 58 36&$-$408&31.6&4.48&1.48&0.91&132& 70.6&{\sc dbb445,ww466}	        \\
79&CHVC 127.1$-$41.4$-$349&01 04 45&21 20 27&$-$347&35.7&3.54&1.34&1.15&178& 63.8&{\sc dbb446}      	        \\
80&CHVC 127.4$-$41.1$-$330&01 05 50&21 38 01&$-$328&35.1&4.34&1.49&0.98& 38& 74.2&{\sc dbb446}      	        \\
81&CHVC 129.3$-$25.2$-$238&01 20 28&37 18 02&$-$238&35.0&0.48&1.10&0.89&  5&  5.5&            	        \\
82&CHVC 129.8$-$16.9$-$255&01 28 47&45 29 16&$-$256&44.5&0.76&0.97&0.78&166&  6.7&            	        \\
83&CHVC 132.2$-$36.7$-$345&01 24 30&25 33 44&$-$343&28.3&1.62&0.88&0.80& 20& 13.3&            	        \\
84&CHVC 132.4$-$17.1$-$288&01 42 27&44 51 43&$-$289&40.1&0.81&0.88&0.78&176&  6.5&            	        \\
85&CHVC 134.0$-$37.6$-$261&01 30 04&24 27 08&$-$259&32.1&0.93&1.15&1.11& 73& 13.9&            	        \\
86& HVC 134.5$-$31.9$-$373&01 36 41&29 58 57&$-$371&36.8&2.90&1.96&1.25& 79& 83.2&           		        \\
87&CHVC 138.0$-$31.7$-$328&01 50 24&29 24 43&$-$325&34.6&0.98&1.21&0.82&145& 11.4&            	        \\
88&CHVC 138.4$-$32.7$-$321&01 50 26&28 25 27&$-$318&35.3&1.27&1.56&1.04&171& 24.1&            	        \\
89&CHVC 139.4$-$38.5$-$304&01 47 05&22 37 48&$-$300&26.9&1.25&1.13&0.78& 21& 12.9&            	        \\
90&CHVC 141.8$-$38.3$-$311&01 55 15&22 15 29&$-$307&24.8&0.58&0.84&0.78& 90&  4.4&            	        \\
91&CHVC 144.3$-$30.5$-$169&02 15 22&28 49 00&$-$165&34.7&1.61&1.23&0.92& 33& 21.3&            	        \\
92& HVC 147.8$-$31.9$-$146&02 25 30&26 26 15&$-$141&39.3&9.30&1.06&0.87&122&100.4&{\sc dbb515,ww471}	        \\
93&CHVC 149.1$-$35.0$-$245&02 24 02&23 09 11&$-$240&30.4&0.88&0.87&0.78& 41&  7.0&            	        \\
94&CHVC 154.0$-$36.4$-$256&02 36 24&20 01 05&$-$250&33.5&0.83&1.44&0.78& 64& 10.9&            	        \\
95&CHVC 154.8$-$30.9$-$270&02 50 55&24 31 14&$-$264&36.3&0.61&0.84&0.78& 90&  4.7&            	        \\

\hline
\end{tabular}}
\begin{list}{}{}
\item[$^{\mathrm{a}}$] Catalog ID's prefaced with DBB refer to the entries
of Table~1 of De Heij et al. \cite{dehe02a}, those prefaced with BB to Table~1
of Braun \& Burton \cite{brau99} and those with WW to Table~1 of Wakker \& 
van Woerden \cite{wakk91}.
\end{list}
\end{table*}

\begin{acknowledgements}
We thank Jay Lockman for obtaining the confirming GBT spectrum of the M31/M33
filament, Frank Briggs for useful discussions concerning normalization
methods of \hi\ distribution functions and Mary Putman for making
a digital version of the HIPASS image of the Magellanic Clouds and
Stream available to us. The Westerbork Synthesis Radio
Telescope is operated by the Netherlands Foundation for Research in
Astronomy under contract with the Netherlands Organization for
Scientific Research.
\end{acknowledgements}

\end{document}